\documentclass[aps,prd,reprint,preprintnumbers,amsmath,amssymb,floatfix,superscriptaddress]{revtex4-1}

\usepackage{graphicx}
\usepackage{epsfig}
\usepackage{dcolumn}
\usepackage{bbm}
\usepackage[colorlinks,linkcolor=red,anchorcolor=green,citecolor=blue,CJKbookmarks=True]{hyperref}
\usepackage{braket}

\usepackage{multirow} 

\begin{document}
\title{Decays of $1^{-+}$ charmoniumlike hybrid}

\author{\small Chunjiang Shi}
\email{shichunjiang@ihep.ac.cn}
\affiliation{Institute of High Energy Physics, Chinese Academy of Sciences, Beijing 100049, People's Republic of China}
\affiliation{School of Physical Sciences, University of Chinese Academy of Sciences, Beijing 100049, People's Republic of China}

\author{\small Ying Chen}
\email{cheny@ihep.ac.cn}
\affiliation{Institute of High Energy Physics, Chinese Academy of Sciences, Beijing 100049, People's Republic of China}
\affiliation{School of Physical Sciences, University of Chinese Academy of Sciences, Beijing 100049, People's Republic of China}

\author{\small Ming Gong}
\affiliation{Institute of High Energy Physics, Chinese Academy of Sciences, Beijing 100049, People's Republic of China}
\affiliation{School of Physical Sciences, University of Chinese Academy of Sciences, Beijing 100049, People's Republic of China}

\author{\small Xiangyu Jiang}
\affiliation{Institute of High Energy Physics, Chinese Academy of Sciences, Beijing 100049, People's Republic of China}
\affiliation{School of Physical Sciences, University of Chinese Academy of Sciences, Beijing 100049, People's Republic of China}

\author{\small Zhaofeng Liu}
\affiliation{Institute of High Energy Physics, Chinese Academy of Sciences, Beijing 100049, People's Republic of China}
\affiliation{School of Physical Sciences, University of Chinese Academy of Sciences, Beijing 100049, People's Republic of China}
\affiliation{Center for High Energy Physics, Peking University, Beijing 100871, People's Republic of China}

\author{\small Wei Sun}
\affiliation{Institute of High Energy Physics, Chinese Academy of Sciences, Beijing 100049, People's Republic of China}

\def\modified#1{\red{#1}}
\begin{abstract}
By extracting the transition amplitudes, we give the first lattice QCD prediction of the two-body decay partial widths of the $1^{-+}$ charmoniumlike hybrid $\eta_{c1}$. Given the calculated mass value $m_{\eta_{c1}}=4.329(36)$ GeV, the $\eta_{c1}$ decay is dominated by the open charm modes $D_1\bar{D}$, $D^*\bar{D}$ and $D^*\bar{D}^*$ with partial widths of $258(133)$ MeV, $88(18)$ MeV and $150(118)$ MeV, respectively. The coupling of $\eta_{c1}$ to $\chi_{c1}$ plus a flavor singlet pseudoscalar is not small, but $\chi_{c1}\eta$ decay is suppressed by the small $\eta-\eta'$ mixing angle. The partial width of $\eta_{c1}\to \eta_c\eta'$ is estimated to be around 1 MeV. We suggest experiments to search for $\eta_{c1}$ in the $P$-wave $D^*\bar{D}$ and $D^*\bar{D}^*$ systems. Especially, the polarization of $D^*\bar{D}^*$ can be used to distinguish the $1^{-+}$ product (total spin $S=1$) from $1^{--}$ products ($S=0$).

\end{abstract}
\maketitle
\section{Introduction}
Quantum Chromodynamics (QCD) permits the existence of hybrid mesons that are composed of both quarks and gluons, among which the $J^{PC}=1^{-+}$ states are most intriguing since this quantum number is prohibited for $q\bar{q}$ states. Up to now there are three experimental candidates for $I^G J^{PC}=1^-1^{-+}$ light hybrid mesons, namely, $\pi_1(1400)$~\cite{IHEP-Brussels-LosAlamos-AnnecyLAPP:1988iqi}, $\pi_1(1600)$~\cite{E852:1998mbq,COMPASS:2018uzl,JPAC:2018zyd} and $\pi_1(2105)$~\cite{E852:1998mbq} (details can be found in the latest review~\cite{Chen:2022asf} and the references therein). Recently, the BESIII collaboration reported the first observation of a $I^GJ^{PC}=0^+1^{-+}$ structure $\eta_1(1855)$ in the $\eta\eta'$ system through the partial wave analysis of the $J/\psi\to \gamma \eta\eta'$ process~\cite{BESIII:2022riz,BESIII:2022iwi}, which is a candidate for light isoscalar $1^{-+}$ hybrid~\cite{Chen:2022qpd,Qiu:2022ktc,Shastry:2022mhk}. It is natural to anticipate the existence of the $1^{-+}$ charmoniumlike hybrid (named $\eta_{c1}$, whose mass is predicted to be $m_{\eta_{c1}}\sim 4.2-4.4$ GeV by lattice QCD studies~\cite{Liao:2002rj,Mei:2002ip,Dudek:2008sz,Dudek:2009kk,HadronSpectrum:2012gic,Yang:2012gz,Ma:2019hsm,Brambilla:2019esw}. In order to identify $\eta_{c1}$ experimentally, its decay properties is also very demanding. Although the flux-tube model gives a selection rule that a hybrid meson decays preferably into final states of an $S$-wave meson and a $P$-wave meson~\cite{Isgur:1984bm,Close:1994hc,Page:1996rj,Page:1998gz}, we would like to perform an \textit{ab initio} investigation of hybrid decays through the lattice QCD approach. 

The state-of-art lattice QCD approach to study strong decays of hadrons is the L\"uscher method~\cite{Luscher:1986pf,Luscher:1990ux,Luscher:1991cf} and its generalization that takes the coupled channel effects into account (see the review articles Ref.~\cite{Briceno:2017max,Mai:2022eur} and the references therein).
For $\eta_{c1}$ with a mass as high as roughly 4.3 GeV, many open-charm and closed-charm decay channels are opened. 
In order to tackle the complicated coupled channel effects, the related study using the (generalized) L\"uscher method requires quite a lot of finite volume energy levels to be determined as precisely as possible. The calculation should be carried out on multiple lattice volumes and in different moving frames. This is numerically and computationally very challenging. The computational complication of this sort of studies can be learnt by the recent lattice QCD studies by the Hadron Spectrum Collaboration on the strong decays of the light hybrid $\pi_1$~\cite{Woss:2020ayi} and on the charmonium-like systems~\cite{Wilson:2023anv,Wilson:2023hzu}.

In order to take a quick look at the decay properties of $\eta_{c1}$, we adopt the method proposed by Michael and McNeile (M\&M)~\cite{McNeile:2002az,McNeile:2002fh} to calculate the tree-level transition amplitudes for $\eta_{c1}$ decays, from which the effective couplings can be estimated. Previously lattice studies using the M\&M method gave reasonable results for the decays of some mesons~\cite{McNeile:2004rf,Michael:2005kw,McNeile:2006bz,McNeile:2006nv,Hart:2006ps,Michael:2006hf}. The M\&M method is also applied to the study of the decay process $\Delta\to N\pi$~\cite{Alexandrou:2013ata,Alexandrou:2015hxa} and the results are consistent with the those from the L\"{u}scher method~\cite{Andersen:2017una,Silvi:2021uya,Morningstar:2021ewk} and physical values~\cite{Pascalutsa:2005vq,Hemmert:1994ky}. 
In Ref.~\cite{Bali:2015gji}, the M\&M method is applied to the decay process $\rho\to\pi\pi$ and obtains the effective coupling constants $g_{\rho\pi\pi}$ ranging from 5.2 to 8.4 (from different lattice volumes and different $\pi\pi$ kinetic configurations), which  is compatible with the value $g_{\rho\pi\pi}\sim 6.0$ that is derived by the width of $\rho$ meson~\cite{ParticleDataGroup:2022pth} and is determined by lattice QCD studies using the L\"uscher method~\cite{CP-PACS:2007wro,Budapest-Marseille-Wuppertal:2010gis,Lang:2011mn,CS:2011vqf,Pelissier:2012pi,Dudek:2012xn,Fahy:2014jxa,Wilson:2015dqa,Bali:2015gji,Bulava:2015qjz,Guo:2015dde,Fu:2016itp,Hu:2016shf,Alexandrou:2017mpi,Hu:2017wli,Andersen:2018mau,Erben:2019nmx,ExtendedTwistedMass:2019omo,Fischer:2020yvw,Akahoshi:2021sxc,Mai:2022eur}. Obviously, the discrepancy of $g_{\rho\pi\pi}$ obtained by the M\&M method from the physical value can be as large as 40-50\%, and serves as an estimate of the systematic uncertainty of the M\&M method. 
With this sizeable uncertainty in mind, we think the results obtained from the M\&M method are helpful and informative for us to learn qualitatively the decay properties of $\eta_{c1}$ from first principles. 

This article is organized as follows. We introduce the theoretical formalism of this study in Sect.~\ref{sec:formalism}, which includes the explicit expressions of the effective Lagrangian for the $\eta_{c1}$ decays and the basic logic of the M\&M method. The numerical procedures and results are presented in Sec.~\ref{sec:numerical} (for the convenience of the readers who are interested only in the physical results, we leave the very details of the lattice treatments, such as the operator construction, the derivation of the decay amplitudes and the data analysis, to the Appendix part), Section~\ref{sec:discussion} is devoted to the discussions and the physical implications of our results. Sect.~\ref{sec:summary} is the summary of this work. 

\section{Formalism}\label{sec:formalism}
The two-body decay $\eta_{c1}\to AB$ includes closed-charm decays $\chi_{c1}\eta(\eta')$, $\eta_c \eta(\eta')$, $J/\psi\omega(\phi)$, and open-charm modes $D_{(s)}^*\bar{D}_{(s)}$, $D_{(s)}^*\bar{D}_{(s)}^*$, $D\bar{D}_1(2420)$, etc. The effective Lagrangian for closed-charm decays can be written as 
\begin{equation}\label{eq:L1}
    \begin{aligned}
        \mathcal{L}_\mathrm{I}^\mathrm{cc}\sim& -g_{\chi\eta}m_{\eta_{c1}} H_\mu A^\mu \eta-ig_{\eta_c\eta} H_\mu \eta_c \overleftrightarrow{\partial}^\mu \eta\\
        &+i H_\mu \left(g\psi_\nu\partial^\nu \omega^{\mu}+g'\omega_\nu \partial^\nu \psi^\mu+g_0\psi_\nu \overleftrightarrow{\partial}^\mu \omega^{\nu}\right),     
    \end{aligned}
\end{equation}
where $\overleftrightarrow{\partial}$ represents $\overleftarrow{\partial}-\overrightarrow{\partial}$, and the fields $H_\mu$, $\chi_\mu$, $\eta_c$, $\eta$, $\psi_\mu$ and $\omega_\mu$ are for $\eta_{c1}$, $\chi_{c1}$, $\eta_c$, $\eta(\eta')$, $J/\psi$ and $\omega$ mesons, respectively. 

For open-charm decays, the quantum numbers $I^GJ^{PC}=0^+1^{-+}$ of $\eta_{c1}$ constrain the flavor structure of the $D\bar{D'}$($D'$ here refers to either $D^*$ and $D_1$) final state completely. We take the conventions $\mathcal{C} | D \rangle = + |\bar{D}\rangle$, $\mathcal{C} | D^* \rangle = - |\bar{D}^* \rangle$, $\mathcal{C} | D_1 \rangle = + |\bar{D}_1 \rangle$ for the charge conjugate transformation ($\mathcal{C}$) of charm mesons. 
When the isospin doublet fields $D_{1\mu}$, $D_\mu$ and $D$ are introduced for $D_1$, $D^*$ and $D$ mesons of the flavor wave functions $(|c\bar{d}\rangle,-|c\bar{u}\rangle)^T$, the effective Lagrangian for open-charm decays reads
\begin{equation}\label{eq:L2}
\begin{aligned}
     \mathcal{L}_\mathrm{I}^\mathrm{oc}&\sim  g_{D_1D} m_{\eta_{c1}} H_\mu\frac{1}{2}\left(D_1^{\mu,\dagger} D+D^\dagger D_1^\mu\right)\\
     &+g_{D^*\bar{D}^*} H^\mu \frac{i}{\sqrt{2}}\left(D^{\nu,\dagger}\partial_\nu D_\mu+\partial_\nu D^{\dagger}_\mu D^\nu\right)\\
     &+\frac{g_{D^*\bar{D}}}{m_{\eta_{c1}}} \epsilon^{\mu\nu\rho\sigma} (\partial_\mu H_\nu) \frac{1}{2}
     \left[(\partial_\rho D_{\sigma}^{\dagger}) D- D^\dagger(\partial_\rho D_\sigma)\right].
\end{aligned}
\end{equation}
So the prediction of partial decay widths depends on a reliable determination of the effective couplings in the effective Lagrangian. 

According to the effective Lagrangian Eq.~(\ref{eq:L1}) and Eq.~(\ref{eq:L2}), the tree-level transition amplitudes
 $   x_{AB}^{(\lambda'\lambda'')\lambda}\approx \langle AB;(\lambda'\lambda''),\vec{k}|\hat{H}|\eta_{c1}, \lambda \rangle$
 in the rest frame of $\eta_{c1}$ are expressed explicitly as
\begin{equation}\label{eq:amplitude}
    \begin{aligned}
        x_{AP}^{\lambda'\lambda}=& {g}_{AP} m_{\eta_{c1}} \vec{\epsilon}_\lambda(\vec{0})\cdot \vec{{\epsilon}}^{~*}_{\lambda'}(\vec{k}),\\
        x_{PP}^\lambda=& 2{g}_{PP} \vec{{\epsilon}}_\lambda(\vec{0})\cdot \vec{k},\\
        x_{D^*\bar{D}}^{\lambda'\lambda}=& {g}_{D^*\bar{D}} \vec{\epsilon}_\lambda(\vec{0})\cdot(\vec{\epsilon}^{~*}_{\lambda'}(\vec{k})\times \vec{k}), \\
        x_{D^*\bar{D}^*}^{\lambda'\lambda''\lambda}=& 2{g}_{D^*\bar{D}^*}\vec{\epsilon}_\lambda(\vec{0})\cdot\left(\vec{k}\times \left[\vec{\epsilon}^{~*}_{\lambda'}(\vec{k})\times \vec{\epsilon}^{~*}_{\lambda''}(-\vec{k})\right]\right),
    \end{aligned}
\end{equation}
where $\epsilon^\mu_\lambda(\vec{0})$,  $\epsilon_{\lambda'}(\vec{k})$ and $\epsilon_{\lambda''}(-\vec{k})$ are the polarization vectors 
of $\eta_{c1}$, $A$ (if a (an axial) vector) and $B$ (if a (an axial) vector). The $J/\psi\omega$ decay is strongly suppressed due to the OZI rule and will not be discussed in depth in this Letter, so the explicit expression of $x_{J/\psi\omega}$ is omitted here.

On the lattice, the transition amplitude $x_{AB}$ for $\eta_{c1}\to AB$ appears in the temporal transfer matrix $\hat{T}=e^{-a_t \hat{H}}$, where $a_t$ is the lattice spacing in the time direction. If the coupled channel effect is tentatively ignored, $\hat{T}$ is parameterized as 
\begin{equation}
    \hat{T}=e^{-a_t\hat{H}}=e^{-a_t\bar{E}} \left(
    \begin{array}{cc}
    e^{-a_t\Delta/2} & a_t x_{AB}\\
    a_t x_{AB} & e^{+a_t\Delta/2}
    \end{array}\right)
\end{equation}
in the Hilbert space spanned by $|\eta_{c1}\rangle$ and $|AB\rangle$ (normalized as $\langle\eta_{c1}|\eta_{c1}\rangle=1$ and $\langle{AB}|{AB}\rangle=1$), where $\bar{E}=(m_{\eta_{c1}}+E_{AB})/2$ and $\Delta=m_{\eta_{c1}}-E_{AB}$ have been defined with $E_{AB}=E_A+E_B$. It is easy to show that, when each single particle state $|X\rangle$ is normalized as $\langle X|X\rangle=2E_X L^3$ with $L^3$ being the spatial volume of the lattice and $E_X$ being the energy of $|X\rangle$, then $x_{AB}^{(\lambda'\lambda'')\lambda}$ is encoded in the following ratio function~\cite{McNeile:2004rf,Michael:2005kw,Alexandrou:2015hxa},
\begin{eqnarray}\label{eq:ratio}
    R_{AB}(\vec{k},t)=&\frac{C^{33}_{AB,\eta_{c1}}(\vec{k},t)}{\sqrt{[C_{AA}(\vec{k},t)C_{BB}(-\vec{k},t)]^{33} C^{33}_{\eta_{c1},\eta_{c1}}(\vec{0},t)}}\nonumber\\
        \approx& t\left(1+\frac{1}{24}(a_t\Delta t)^2\right)\frac{1}{\sqrt{8L^3 m_{\eta_{c1}}E_A E_B}}\nonumber\\
        \times&\sum\limits_{\lambda(\lambda'\lambda'')} \frac{a_t x^{(\lambda'\lambda'')\lambda}_{AB}[\epsilon_{(\lambda'\lambda'')}^3(\vec{k},AB)\epsilon^{3,*}_\lambda(\vec{0})]}{\sqrt{\mathcal{P}_{A}(\vec{k}) \mathcal{P}_{B}(-\vec{k}) \mathcal{P}_{\eta_{c1}}^{33}(\vec{0})}},
\end{eqnarray}
for $a_t\Delta\ll 1$, where $C^{ii}_{AB,\eta_{c1}}(\vec{k},t)$ is the correlation function of operators $\mathcal{O}^i_{\eta_{c1}}$ and $\mathcal{O}^i_{AB}(\vec{k})$ that create (or annihilate) the states $\eta_{c1}$ and $A(\vec{k})B(-\vec{k})$ respectively, and $C_{XX}(\vec{p},t)$ is the correlation function for $X$ moving with a momentum $\vec{p}$. Here $X$ refers to $\eta_{c1}$, $A$ or $B$. The polarization of the $AB$ state, $\epsilon_{(\lambda'\lambda'')}^\mu(\vec{k},AB)$, depends on the wave functions of $A$ (either 1 or $\epsilon_{\lambda'}(\vec{k})$ ) and $B$ (either 1 or $\epsilon_{\lambda''}(-\vec{k})$), as well as the relative momentum $\vec{k}$ (see Appendix.\ref{sec:appendix:parametrization-correlation} for the explicit expressions). The quantity $\mathcal{P}_X(\vec{k})$ comes from the completeness of polarization vectors of $X$ and takes the value of unity for a pseudoscalar and $\mathcal{P}_X^{ij}=\delta^{ij}+\frac{k^i k^j}{m_X^2}$ for a vector. Note that the major assumptions in the derivation of Eq.~(\ref{eq:ratio}) are 
\begin{equation}\label{eq:approx}
   \langle 0|\mathcal{O}_{\eta_{c1}}|AB\rangle\approx 0 ~~\langle 0|\mathcal{O}_{AB}| \eta_{c1} \rangle\approx 0.
\end{equation}
This can be checked by looking at the deviation of $R_{AB}(\vec{0},t=0)$ from zero (see Appdenix D).

Obviously, $R_{AB}(t)$ in Eq.~(\ref{eq:ratio}) is linear in the time range where excited state contamination is negligible and $a_t\Delta t$ is sufficiently small. This allows the transition matrix $a_t x_{AB}^{(\lambda'\lambda'')\lambda}$, and thereby the effective coupling $g_{AB}$, to be extracted from the slope of $R_{AB}(t)$.

\begin{table}[t]
    \renewcommand\arraystretch{1.5}
    \caption{Parameters of the gauge ensembles. $N_V$ is the number of the eigenvectors that span the Laplacian subspace~\cite{Peardon:2009gh}.}
    \label{tab:config}
    \begin{ruledtabular}
        \begin{tabular}{lccccccc}
            IE & $N_s^3\times N_t$     & $\beta$ & $a_t^{-1}$(GeV) & $\xi$      & $m_\pi$(MeV) &  $N_V$  & $N_\mathrm{cfg}$ \\\hline
            L16 & $16^3 \times 128$ & 2.0     & $6.894(51)$     & $\sim 5.3$ & $\sim 350$       &  70     & $708$           \\
            L24 & $24^3 \times 192$ & 2.0     & $6.894(51)$     & $\sim 5.3$ & $\sim 350$       &  160    & $171$\\
        \end{tabular}
    \end{ruledtabular}
\end{table}
\begin{table*}[tbh!]
    \renewcommand\arraystretch{1.5}
    \caption{The masses of light hadrons, charmed mesons and charmonium on the two lattices. The PDG value of $m_{D^{(*)}}$ is the average of masses of $D^{(*)0}$ and $D^{(*)+}$.}
    \label{tab:spectrum}
    \begin{ruledtabular}
        \begin{tabular}{lllllllll|c}
        $X$             &  $\omega$  & $\eta_{(2)}$ & $\eta_c$  & $J/\psi$  & $\chi_{c1}$ & $D$       &  ${D^*}$      &  $D_1$        & $\eta_{c1}$\\\hline
        $m_X$(L16)(MeV) &  857(10)    & 715(10)   & 2975.0(3)         & 3099.1(2)         & 3559.8(1.7) & 1882(1)   & 2023(1)       &  2423(7)      & 4330(21)  \\
        $m_X$(L24)(MeV) &  845(7)    & 705(21)  & 2976.3(4)         & 3099.4(4)         & 3560.4(6.4) & 1881(1)   & 2019(1)       &  2430(10)     & 4328(68)  \\
        $m_X$(PDG)(MeV)~\cite{ParticleDataGroup:2022pth}&  782       & 547/958    & 2983      & 3097      & 3511         & $\sim 1867$ & $\sim 2008$ &  2420   &  \--     \\
        \end{tabular}
    \end{ruledtabular}
\end{table*}
\section{Numerical details}\label{sec:numerical}
The calculations are performed on two $N_f=2$ gauge ensembles at the same pion mass $m_\pi\approx 350$ MeV, the same temporal lattice spacing $a_t$, but with different lattice sizes, namely, $N_s^3\times N_t=16^3\times 128$  and $24^3\times 192$ (labeled as L16 and L24, respectively). 
Both ensembles are generated on anisotropic lattice with the same anisotropic parameter $\xi=a_s/a_t=5.3$ ($a_s$ is the lattice spacing in spatial directions)~\cite{Jiang:2022ffl}. 
The parameters of the gauge ensembles are collected in Table~\ref{tab:config}.
We adopt the tadpole improved gauge action~\cite{Morningstar:1997ff,Chen:2005mg} for gluons and the tadpole improved anisotropic clover fermion action in Ref.~\cite{Zhang:2001in, Su:2004sc, CLQCD:2009nvn} for light quarks. 
The valence charm quark mass parameters are set by $(m_{\eta_c}+3m_{J/\psi})/4=3069$ MeV.
Within the framework of the distillation method, the perambulators of light $u,d$ quarks and charm quark are calculated with the time sources running over all the time slices and the number of eigenvectors spanning the Laplacian subspace are chosen to be $N_V=70$ for L16 and $N_V=160$ for L24. Since all the mesons involved in $\eta_{c1}$ decays are the ground states in specific $J^{PC}$ channels, and the distillation method provides an inherent smearing scheme for quark fields, we use only the simplest generic quark bilinear operators for all the mesons. 
Table~\ref{tab:spectrum} lists the masses of mesons involved in this calculation. Since we work in the $N_f=2$ QCD, we label the flavor singlet (isoscalar) pseudoscalar 
meson by $\eta_{(2)}$ to differentiate it from the physical states $\eta$ and $\eta'$ when we present the corresponding lattice results. We keep the name $\omega$ for the isoscalar vector meson in our lattice setup by considering that the physical $\omega$ meson is nearly a pure $(u\bar{u}+d\bar{d})$ meson.

\begin{figure}[t]
    \centering
    \includegraphics[width=0.9\linewidth]{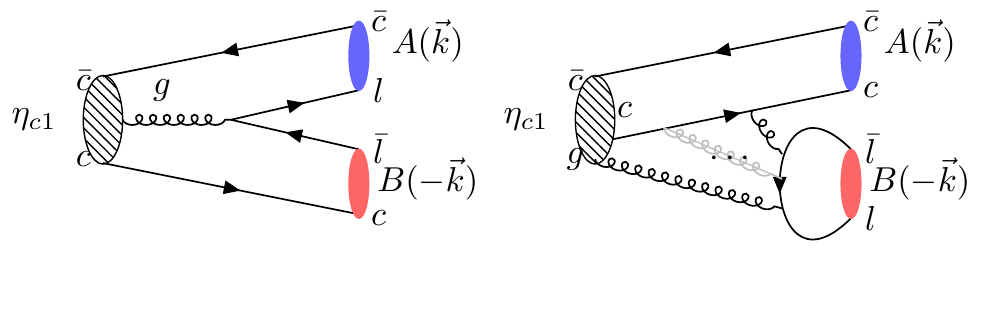}
    \caption{Schematic diagrams illustrating the decay process $\eta_{c1}\to AB$. 
    The left panel depicts the open-charm decay, while the right panel is for the closed-charm decay (at least two intermediate gluons for a light pseudoscalar and three intermediate
    gluons for a light vector meson in the $AB$ system).}
    \label{fig:quark-diagram}
\end{figure}

The key task is to calculate the correlation function $C^{ii}_{AB,\eta_{c1}}(\vec{k},t)$, where both operators $\mathcal{O}^i_{\eta_{c1}}$ and $\mathcal{O}^i_{AB}(\vec{k})$ belong to the $T_1$ representation of the octahedral group $O$ on the lattice.
We use the partial-wave method~\cite{Feng:2010es, Wallace:2015pxa, Prelovsek:2016iyo} to construct $\mathcal{O}^i_{AB}(\vec{k})$.
$\mathcal{O}^i_{\eta_{c1}}$ takes the generic quark bilinear form $\epsilon_{ijk}\bar{c}\gamma_j B_k c$ with $B_k$ being the chromomagnetic field strength, following the operator construction in Ref.~\cite{Basak:2005ir, Dudek:2007wv, Dudek:2009qf, Dudek:2010wm, Thomas:2011rh}.
The operator $\mathcal{O}^i_{AB}(\vec{k})$ is built from the single particle operators $\mathcal{O}_A(\vec{k})$ and $\mathcal{O}_B(-\vec{k})$ through a specific combination $\left[\mathcal{O}_A(\vec{k})\circ\mathcal{O}_{B}(-\vec{k})\right]^i$ (the flavor structure is also considered for open-charm modes).
See the Appendix.\ref{sec:appendix:operators} for the details. 
The schematic quark diagrams for the two-body $\eta_{c1}$ decays are shown in Fig.~\ref{fig:quark-diagram} where the left panel is for the open-charm decays, in which the original gluon in $\eta_{c1}$ splits into a light quark anti-quark pair $l\bar{l}$, which reorganizes with the $c\bar{c}$ pair to form an open-charm final state. 
The right panel depicts the process of the closed-charm decay. Along with the $c\bar{c}$ pair coupling to a final state charmonium, the original gluon and the gluon(s) emitted by the charm quark pair generate a flavor singlet meson (note that the mesons $\eta(\eta'),\omega (\phi)$ have flavor singlet components). 
Obviously, both diagrams have annihilation diagrams of light quarks which can be readily tackled by the distillation method.

\begin{figure}[tbh!]
	\includegraphics[width=1\linewidth]{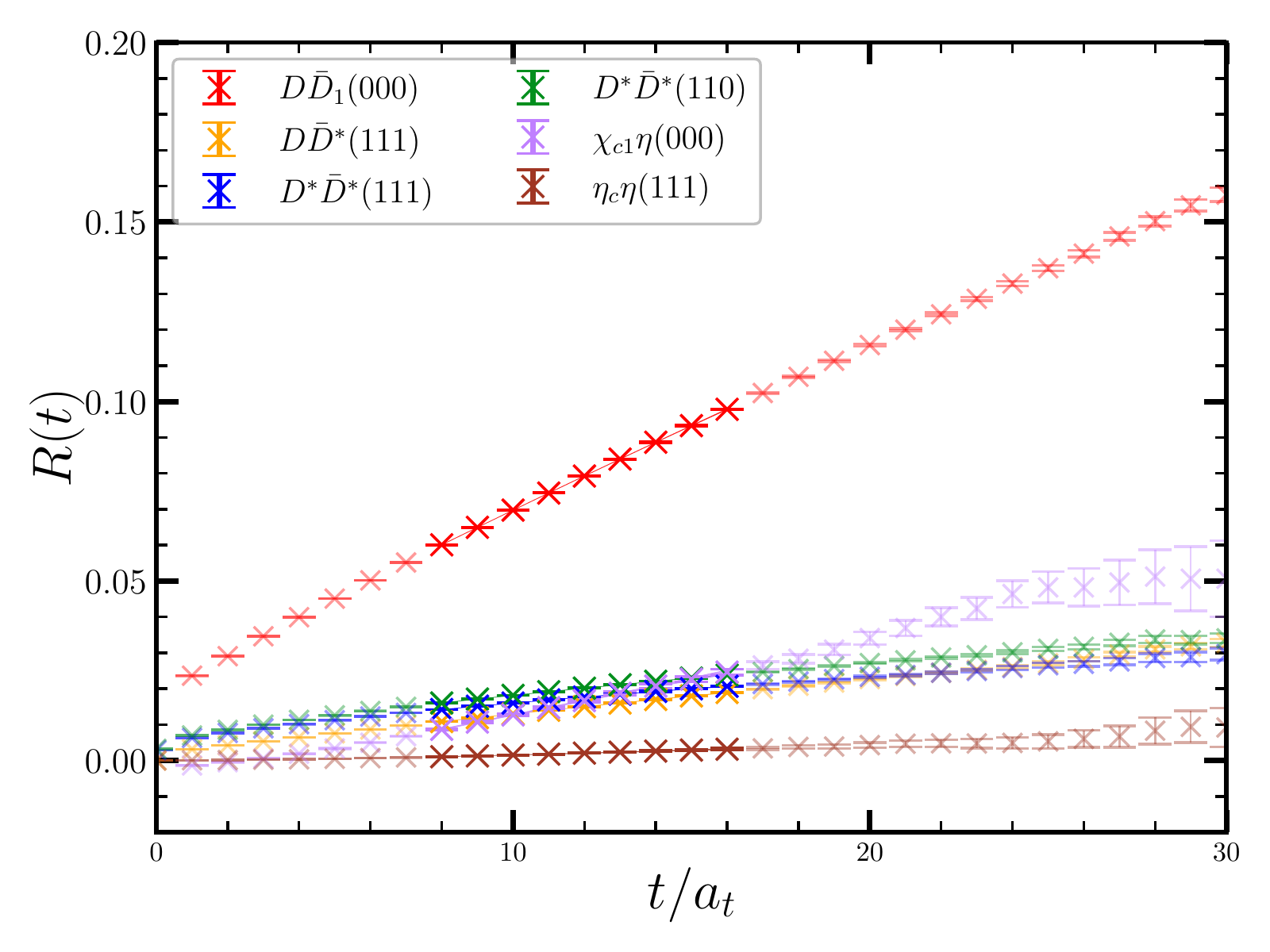}\\
	\includegraphics[width=1\linewidth]{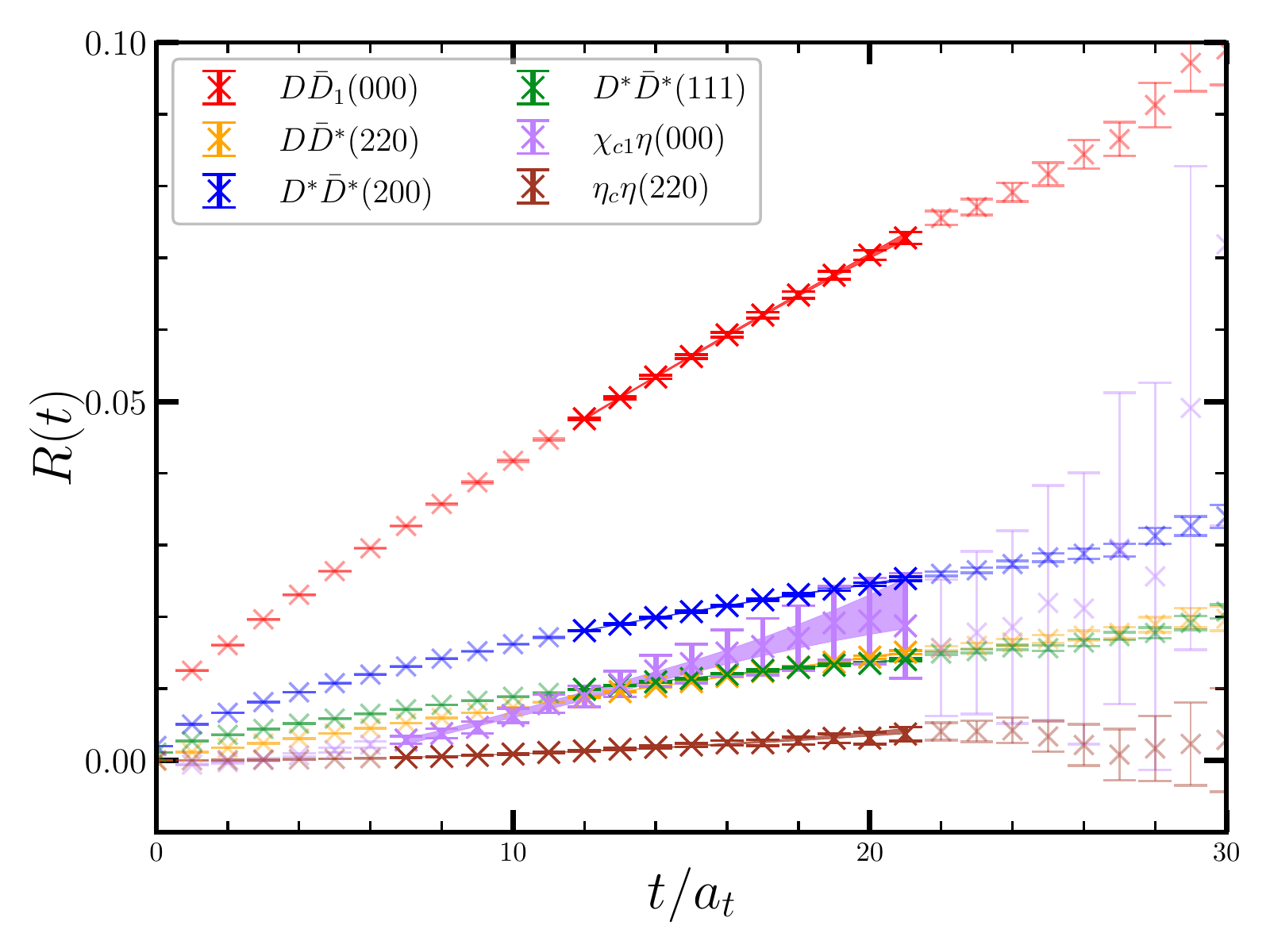}
	\caption{The ratio function $R_{AB}(\vec{k},t)$ on the two gauge ensembles L16 (the upper panel) and L24 (the lower panel) for $\eta_{c1}\to AB$. $AB$ refers to the decay modes $D_1\bar{D}$, $\chi_{c1}\eta_{(2)}$, $D^*\bar{D}$, $D^*\bar{D}^*$, $\eta_c\eta_{(2)}$). The shaded bands illustrate the fit results and the fit window using Eq.~(\ref{eq:ratio-model}).}
	\label{fig:slopes} 
\end{figure}

For each process $\eta_{c1}\to A(\vec{k})B(-\vec{k})$, $R_{AB}(\vec{k},t)$ is easily derived from $C^{ii}_{AB,\eta_{c1}}(\vec{k},t)$ and the corresponding correlation functions of single particles. 
In practice, the relative momentum $\vec{k}=2\pi \hat{k}/L$ ($\hat{k}=(n_1,n_2,n_3)$) is chosen so that the two-particle energy $E_{AB}$ is as close as possible to $m_{\eta_{c1}}$, namely $|a_t \Delta|\lesssim 0.01\ll 1$. For some channels, more than one momentum modes $\hat{k}$ are chosen to check the momentum dependence.
Figure~\ref{fig:slopes} shows $R_{AB}(t)$ for all the decay modes concerned on L16 (upper panel) and L24 (lower panel) ensembles. 
It can be seen that
the curves of $R_{AB}(\vec{k},t)$ for open-charm processes and $\chi_{c1}\eta$ exhibit clear linear behavior in the time region $t\in[5,20]$. The curves of $R_{AB}(\vec{k},t)$ for closed-charm decays are indeed linear in $t$ in the same time region but have much smaller slopes, which is certainly the expectation of the OZI rule. 
In the data analysis, we use the following function form to fit the data
\begin{eqnarray}\label{eq:ratio-model}
    R(t)&=&r_0+r_1 t+r_3 t^3.
\end{eqnarray}
 The shaded bands in Fig.~\ref{fig:slopes} illustrate the best fit results in the time windows $[t_\text{min},t_\text{max}]=[8,18]$ for L16 and $[t_\text{min},t_\text{max}]=[7,22]$ for L24 ensembles. The fitted values of $r_0$ for all the channels are few percents or even smaller and therefore justify the approximation of Eq.~(\ref{eq:approx}). 
 The stability of the fit is also checked by fit in different time windows $[t_\text{min},t_\text{min}+10]$ with $t_\text{min}$ varying from 5 to 25.
 It is found that(from Fig.~\ref{fig:fit-stability}), for all the channels, $r_1$ is insensitive to the change of fit range. We take $\delta r_1=(r_1^\text{(max)}-r_1^\text{(min)})/2$ of the fits with $t_\text{min}\in[6,13]$ as a systematic error of $r_1$ and add it to the statistical error in a quadratic way. The final result of $r_1$ for different $AB$ modes are listed in Table~\ref{tab:partial-width}. 

For each decay mode, the connection between $r_1$ and the transition amplitude $x_{AB}$ can be established through Eq.~(\ref{eq:amplitude}) and Eq.~(\ref{eq:ratio}). 
Finally, on the L16 and L24 ensembles, we obtain the effective coupling $g_{AB}$ for each decay mode $AB$ of a specific momentum configuration $\hat{k}$, as shown in Table~\ref{tab:partial-width}. 
For each mode, the values of $g_{AB}$ obtained on the two ensembles L16 and L24 are very close to each other and provide a self-consistent check except for $g_{D^*\bar{D}^*}$. The differences among the values of $g_{D^*\bar{D}^*}$ from L16 and L24 with different momentum configurations of $D^*\bar{D}^*$ is similar to the case of $g_{\rho\pi\pi}$ derived through the M\&M method in Ref.~\cite{Bali:2015gji}. This discrepancy may be due to the possible coupled channel effects and (or) the caveat of the M\&M method itself. In practice, we average the values of $g_{AB}$ to give our final prediction of the effective coupling $g_{AB}$ for each decay mode $AB$, which is also listed in Table~\ref{tab:partial-width} with the error being given by $\delta g_{AB}\sim (g_{AB}^\text{(max)}-g_{AB}^\text{(min)})/2$.

\begin{table}[t]
    \caption{The values of $r_1$ and $g_{AB}$ for different decay modes ($AB$) and different relative momentum modes $\hat{k}=(n_1,n_2,n_3)$ on the two ensembles L16 and L24. The errors of the averaged $\Gamma_{AB}$ is estimated by $\delta g_{AB}\sim (g_{AB}^\text{(max)}-g_{AB}^\text{(min)})/2$.} 
    \label{tab:partial-width}
    \begin{ruledtabular}
     \small
        \begin{tabular}{cccccc}
            Mode           &     $\hat{k}$(IE) & $r_1$    & $g_{AB}$      &$g_{AB}$     & $\Gamma_{AB}$ \\
            ($AB$)         &                   & $(\times 10^{-3})$           &       &    (ave.)    &  (MeV)        \\\hline 
                        &&&&&\\
            \multirow{2}{*}{$D_1\bar{D}$}   &   $(0,0,0)$(L16)     &4.95(5)         &4.27(5)     & \multirow{2}{*}{4.6(6)} & \multirow{2}{*}{258(133)}    \\
                                            &   $(0,0,0)$(L24)     &3.10(26)        &4.92(41)     &                    &     \\
                        &&&&&\\
            \multirow{2}{*}{$D^*\bar{D}$}   &   $(1,1,1)$(L16)     &1.11(3)       &8.35(21)     & \multirow{2}{*}{8.3(7)}& \multirow{2}{*}{88(18)}   \\
                                            &   $(2,2,0)$(L24)     &0.78(7)       &8.34(74)     &  &    \\
                        &&&&\\
            \multirow{4}{*}{$D^*\bar{D}^*$} &   $(1,1,1)$(L16)     &1.00(3)      &3.44(12)      & \multirow{4}{*}{4.6(1.8)}& \multirow{4}{*}{150(118)}   \\
                                            &   $(1,1,0)$(L16)     &1.15(4)      &3.79(12)      &       &      \\
                                            &   $(2,0,0)$(L24)     &1.05(9)      &5.06(42)      &       &    \\
                                            &   $(1,1,1)$(L24)     &0.67(7)      &6.31(58)      &       &    \\
                        &&&&\\
            \multirow{2}{*}{$\chi_{c1}\eta_{(2)}$} &   $(0,0,0)$(L16)           &2.04(26)       &1.31(2)        & \multirow{2}{*}{1.35(45)}   &  \multirow{2}{*}{--}\\
                                                   &   $(0,0,0)$(L24)           &1.18(38)       &1.39(45)       &                           &  \\  
                        &&&&\\
            \multirow{2}{*}{$\eta_c\eta_{(2)}$}     &   $(1,1,1)$(L16)          &0.20(6)        &0.62(18)       & \multirow{2}{*}{0.55(22)}& \multirow{2}{*}{--}\\
                                                    &   $(2,2,0)$(L24)          &0.10(3)        &0.47(12)       &               &  \\

        \end{tabular}
    \end{ruledtabular}
\end{table}

\section{Discussion}\label{sec:discussion}
Although theoretically they are functions of meson masses, the dimensionless $g_{AB}$ are usually assumed to be insensitive to the meson masses in the initial and final states in suitable mass ranges. The physics behind this can be understood as follows. For two-body strong decays of a hadron,
the dictating dynamics is the gluon-quark coupling in QCD. 
Although a rigorous proof is lacking in the low energy regime, it is expected the light quark mass effects on the dimensionless coupling $g_{AB}$ is $\mathcal{O}(m_q/\Lambda_{QCD})$. 
So for $m_q\ll \Lambda_{QCD}$, the quark mass dependence can be neglected when a raw estimate is made.
Consequently, the averaged $g_{AB}$ is applied to predict the partial decay width $\Gamma_{AB}$ of $\eta_{c1}\to AB$ by  
\begin{equation}
\Gamma_{AB}=\frac{1}{8\pi}\frac{k_\text{ex}}{m_{\eta_{c1}}^2}\overline{|\mathcal{M}(\eta_{c1}\to AB)|^2},     
\end{equation}
where $k_\text{ex}$ is the decay momentum determined by the experimental mass values of $A$ and $B$ along with a given $m_{\eta_{c1}}$, $\mathcal{M}$ is the tree-level decay amplitude $x_{AB}$ in Eq.~(\ref{eq:amplitude}) with $\vec{k}$ being replaced by $\vec{k}_\text{ex}$. 
Using the mass $m_{\eta_{c1}}= 4.329(36)$ GeV determined by averaging the values on two ensembles, the partial widths $\Gamma_{AB}$ of all the decay modes concerned are predicted and are shown in Table~\ref{tab:partial-width}. Specifically, the partial widths of open charm decays at $m_{\eta_{c1}}= 4.329(36)$ GeV are 
\begin{eqnarray}
    \Gamma_{D_1\bar{D}}  &=& 258(133) ~\text{MeV},\nonumber\\
    \Gamma_{D^*\bar{D}^*}&=& 150(118)~\text{MeV},\nonumber\\ 
    \Gamma_{D^*\bar{D}}  &=& 88(18)~\text{MeV},
\end{eqnarray}
which indicate $\eta_{c1}$ is a very wide resonance with a total width $\Gamma_{\eta_{c1}}\gtrsim 400\text{MeV}$.
The $D_1\bar{D}$ dominance of the $\eta_{c1}$ decay is in agreement with the expectation of the flux tube model qualitatively, which favors the final state of a $S$-wave meson and a $P$-wave meson a hybrid state. 
However, the partial width in this study is much larger than $\sim 25~\text{MeV}$ predicted by the flux tube model~\cite{Page:1998gz}. 
On the other hand, the $D_1\bar{D}$ mass threshold $E_\text{th}(D_1\bar{D})\approx 4290~\text{MeV}$ is very close to the expected $\eta_{c1}$ mass, so the partial width $\Gamma_{D_1\bar{D}}$ is very sensitive to the $m_{\eta_{c1}}$. There is a possibility that this decay mode is kinetically prohibited if $m_{\eta_{c1}}<E_\text{th}(D_1\bar{D})$.

It is most striking that the ${D^*\bar{D}^*}$ and $D^*\bar{D}$ decays have large branching fractions and even dominate the $\eta_{c1}$ decay when $m_{\eta_{c1}}<E_\text{th}(D\bar{D}_1)$.
The large $\Gamma_{D^*\bar{D}^*}$ and $\Gamma_{D^*\bar{D}}$ refute the expectation of the flux tube model that the decay modes of two $S$-wave mesons are disfavored and that decay modes with two spatially identical particles such as $D^*\bar{D}^*$ is prohibited~\cite{Isgur:1984bm,Close:1994hc,Page:1996rj,Page:1998gz}. We notice that a recent phenomenological study in the Born-Oppenheimer approximation also indicates the non-suppression of the $D^{(*)}\bar{D}^{(*)}$ decay modes of the $1^{-+}$ heavy quarkonium-like hybrids~\cite{Bruschini:2023tmm}.
Our observation suggests $D^*\bar{D}$ and $D^*\bar{D}^*$ are good channels for $\eta_{c1}$ to be searched experimentally. 
First, the $D^*\bar{D}$ and $D^*\bar{D}^*$ decay modes are kinetically favored since their thresholds are much lower than the expected $m_{\eta_{c1}}$. 
Secondly, $D$ and $D^*$ are easier to be reconstructed than $D_1$ in experiments. 
One feature of the $D^*\bar{D}^*$ channel is that, its $C$-parity can be determined by measuring the polarizations of $D^*$ and $\bar{D}^*$, since the $C$-parity of a $P$-wave $D^*\bar{D}^*$ system is determined by its total spin $S$ through $C=(-)^{L+S}$. 
This feature can be used to distinguish the possible $1^{-+}$ state from the $1^{--}$ states in the $D^*\bar{D}^*$ system.

For the closed charm decay modes, the tiny partial width of the $J/\psi \omega$ decay is understood due to the OZI rule. 
However, the effective couplings for the $\chi_{c1}\eta_{(2)}$ and $\eta_c\eta_{(2)}$ decays appear not to be particularly small.
This can be attributed to the enhancement by the QCD $U_A(1)$ anomaly~\cite{Jiang:2022gnd,Bali:2014pva}, which introduces an anomalous coupling between gluons and the flavor singlet pseudoscalar meson (the isoscalar $\eta_{(2)}$ in the $N_f=2$ case). 
Since this coupling is proportional to $\sqrt{N_f}$, the coupling $g_{AB}$ for $\chi_{c1}\eta_{(2)}$ and $\eta_{c}\eta_{(2)}$ in Table~\ref{tab:partial-width} should be proportionally increased by a factor of $\sqrt{3/2}$ for the flavor SU(3) case.
Considering that the $\chi_{c1}\eta'$ is likely kinetically prohibited and the $\eta_c\eta$ decay is suppressed by $\sin^2\theta$ with $\theta=-24.5^\circ$ (linear mass relation) or $-11.3^\circ$ (squared mass relation)~\cite{ParticleDataGroup:2022pth}, where $\theta$ is the small $\eta-\eta'$ mixing angle, 
here we restrict our predictions on the partial widths of $\chi_{c1}\eta$ and $\eta_c\eta'$ decays as
\begin{eqnarray}
    \Gamma_{\chi_{c1}\eta}&=& \sin^2\theta\cdot 44(29) ~\text{MeV},\nonumber\\
    \Gamma_{\eta_c\eta'}&=& \cos^2\theta\cdot  0.93(77) ~\text{MeV}.
\end{eqnarray}

\begin{figure}[t]
    \centering
    \includegraphics[width=1.0\linewidth]{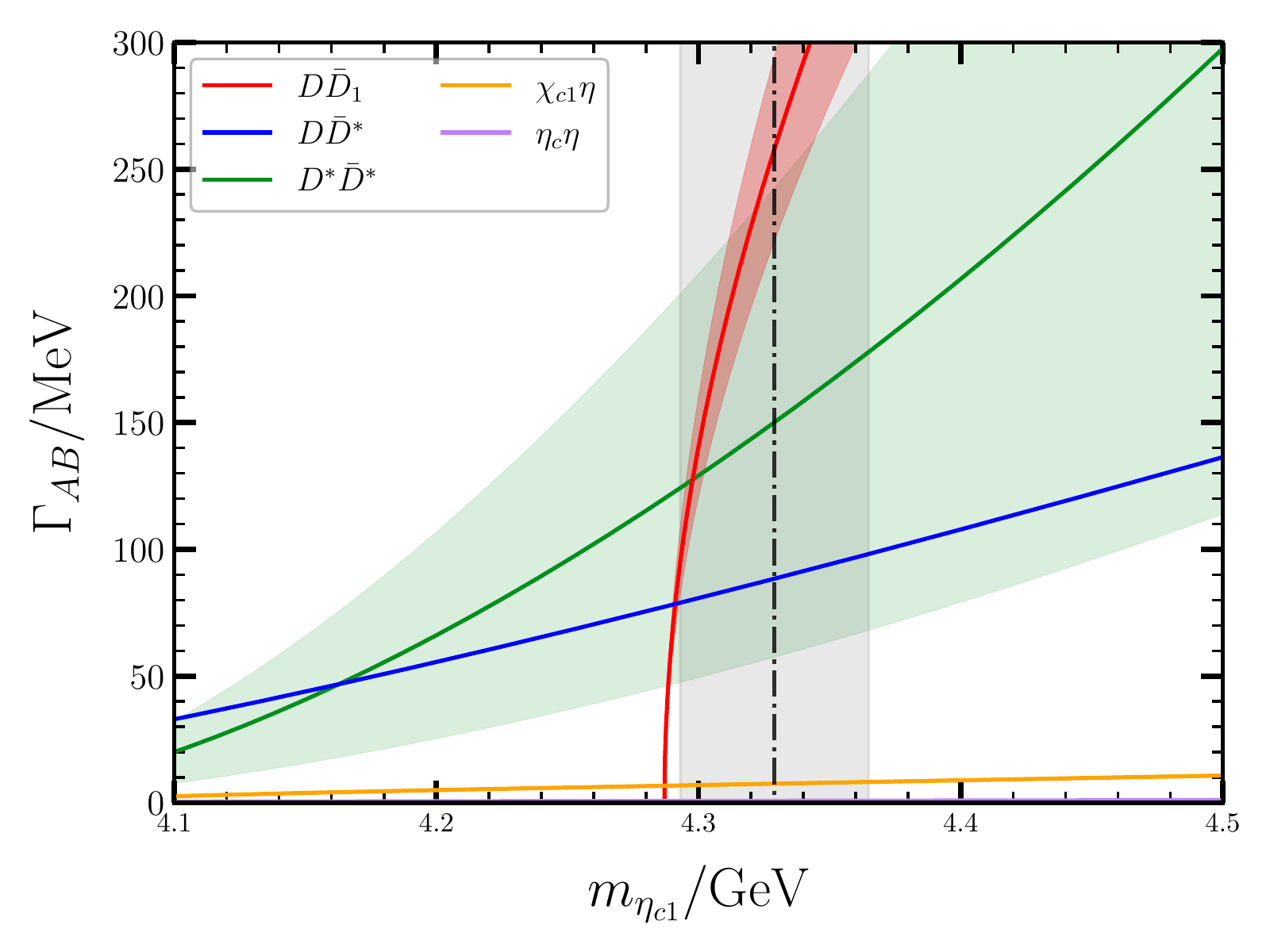}
    \caption{The $m_{\eta_{c1}}$ dependence of the partial decay widths. The vertical dashed line and the grey band illustrate the lattice result of $m_{\eta_{c1}}$ with errors, and the colored bands show the partial decay widths varying with respect to $m_{\eta_{c1}}$ in the range from $4.1$ to $4.5$ GeV. The mixing angle $\theta=-24.5^\circ$ from the linear mass relation~\cite{ParticleDataGroup:2022pth} is used to estimate the partial decay widths of $\chi_{c1}\eta$, $\eta_c\eta'$ decay modes.}
    \label{fig:mass-dependence}
\end{figure}

Although many systematic uncertainties are not well controlled yet, the results in this Letter have already provided very important physical information for $\eta_{c1}$ decay properties. 
Since the open-charm partial widths are sensitive to $m_{\eta_{c1}}$, we plot the $m_{\eta_{c1}}$ dependence of  the partial widths as curves with colored error bands in Fig.~\ref{fig:mass-dependence}, where the vertical grey band illustrates $m_{\eta_{c1}}$ obtained in this work. 
This figure serves as a further guidance for experiments to search for $\eta_{c1}$. In comparison with $e^+e^-$ colliders that favor the production of $1^{--}$ mesons, the $B$ meson decay might be a good place to search for the $\eta_{c1}$ search. 
The LHCb and Belle II experiments may take the mission. 

\section{Summary}\label{sec:summary}
Based on two gauge ensembles of $N_f=2$ dynamical quarks at $m_\pi\approx 350$ MeV, we perform the first lattice QCD calculation of the two-body partial decay widths of the $1^{-+}$ charmonium-like hybrids $\eta_{c1}$. We extract the decay amplitudes for open charm and closed charm decays directly from hadronic correlation functions, and then predict the partial widths $\Gamma_{D_1\bar{D}}=258(133)~\text{MeV}$, $\Gamma_{D^*\bar{D}^*}=150(118)~\text{MeV}$ and $\Gamma_{D^*\bar{D}}=88(18)~\text{MeV}$ for $m_{\eta_{c1}}=4329(36)~\text{MeV}$, which indicate
that $\eta_{c1}$ is a wide resonance. The strikingly large partial widths of $D^*\bar{D}$ and $D^*\bar{D}^*$ decays suggest these system to be good places for the $\eta_{c1}$ search. Especially, the polarization of the $P$-wave $D^*\bar{D}^*$ is an ideal probe to distinguish the $1^{-+}$ products from $1^{--}$ ones. LHCb, Belle II and BESIII experiments may take the mission to search for $\eta_{c1}$ in $D^*\bar{D}$ and $D^*\bar{D}^*$ systems.

\begin{acknowledgments}
We thank Kuang-Ta Chao, Qiang Zhao, Changzheng Yuan, Beijiang Liu and Feng-Kun Guo for valuable discussions. This work is supported by the National Natural Science Foundation of China (NNSFC) under Grants No. 11935017, No. 12293060, No. 12293065, No. 12293061, No. 12075253, No. 12070131001 (CRC 110 by DFG and NNSFC), No. 12175063, No. 12205311. C.Y. also appreciates the support by the National Key Research and Development Program of China (No. 2020YFA0406400), the Strategic Priority Research Program of Chinese Academy of Sciences (No. XDB34030302). The Chroma software system~\cite{Edwards:2004sx} and QUDA library~\cite{Clark:2009wm,Babich:2011np} are acknowledged. The computations were performed on the HPC clusters at Institute of High Energy Physics (Beijing) and China Spallation Neutron Source (Dongguan), and the ORISE computing environment.
\end{acknowledgments}


\section*{Appendix}
\setcounter{equation}{0}
\setcounter{figure}{0}
\setcounter{table}{0}
\renewcommand{\theequation}{A\arabic{equation}}
\renewcommand{\thefigure}{A\arabic{figure}}
\renewcommand{\thetable}{A\arabic{table}}
\renewcommand{\thesection}{}

These appendices provide further information on our study on the partial decay widths of the $1^{-+}$ charmoniumlike hybrid $\eta_{c1}$. We present the details on the parametrization of the correlation function $C^{ij}_{AB,\eta_{c1}}$ and show how the related transition amplitude is encoded. We also give a description of our data analysis in extracting the amplitudes to justify the reliability of the numerical results. 
The procedure of the meson-meson operators with the specific quantum numbers $I^G J^{PC}$ follows the established method, so we put it in the first part for further reference. 

\subsection{Meson-meson operators}\label{sec:appendix:operators}
We use the partial-wave method to construct the interpolating meson-meson (labeled as $A$ and $B$) operators for the specific quantum numbers $J^P$ ~\cite{Feng:2010es, Wallace:2015pxa, Prelovsek:2016iyo}. In general, let $\mathcal{O}_{X}^{M_{X}}(\vec{k})$ be the operator for the particle $X=A$ or $B$ with spin $S_X$ and spin projection $M_{X}$ in the $z$-direction, then for the total angular momentum $J$ and the $z$-axis projection $M$, the relative orbital angular momentum $L$, the total spin $S$, the explicit construction of the $AB$ operator is expressed as 
\begin{widetext}
\begin{equation}\label{eq:two-meson-sm}
\begin{aligned}
    \mathcal{O}_{AB;JLSP}^{M}(\hat{k}) = & \sum\limits_{M_L, M_S, M_{A}, M_{B}} \langle L, M_L; S,M_S| JM \rangle
                                    \langle S_AM_{A}; S_BM_{B}| S, M_S \rangle\\
                            & \times \sum\limits_{R\in O_h} Y^*_{LM_L}(R\circ\vec{k}) \mathcal{O}_A^{M_{A}}(R\circ \vec{k}) \mathcal{O}_B^{M_{B}}(-R\circ \vec{k}),
\end{aligned}
\end{equation}
\end{widetext}
where $\hat{k}=(n_1,n_2,n_3)$ is the momentum mode of $\vec{k}=\frac{2\pi}{La_s}\hat{k}$ with $n_1\ge n_2\ge n_3\ge 0$ by convention, $R\circ\vec{k}$ is the spatial momentum rotated from $\vec{k}$ by $R\in O_h$ with $O_h$ being the lattice symmetry group, $|S,M_S\rangle$ is the total spin state of the two particles involved, $|LM_L\rangle$ is the relative orbital angular momentum state, and $|JM\rangle$ is the total angular momentum state, and $ Y^*_{LM_L}(R\circ\vec{k})$ is the spherical harmonic function of the direction of $R\circ \vec{k}$. 

For the case of this study, $\eta_{c1}$ has quantum numbers $I^G J^{PC}=0^+ 1^{-+}$. So we need the two-meson operators of $J^P=1^-$, which can be deduced from the $T_1^{P}$ representation of $O_h$. The two-body decays of $\eta_{c1}$ include the $S$-wave decay $AP$ (one axial vector meson and one pseudoscalar meson), the $P$-wave $PP$ mode (two pseudoscalars), the $P$-wave $VP$ mode (one vector and one pseudoscalar), and the $P$-wave $VV$ mode. For closed-charm decays, such as $\chi_{c1}\eta(\eta')$ ($AP$), $\eta_c\eta(\eta')$ and $J/\psi \omega(\phi)$, the $\mathcal{C}$-parity of the two meson system is definitely the product of those of the constituent mesons. For open-charm decays such as $D_1\bar{D}$, $D^*\bar{D}$, the flavor structure of the $D\bar{D'}$($D'$ here refers to either $D^*$ and $D_1$) final state should be
\begin{eqnarray}\label{eq:DDstar-opt-sm}
|D\bar{D}'\rangle_{(C=+)}^{(I=0)}&=&\frac{1}{2}\left(|D^+D^{'-}\rangle+|D^0\bar{D}^{'0}\rangle\right)\nonumber\\
&\pm&\frac{1}{2}\left(|D^-D^{'+}\rangle+|\bar{D}^0D^{'0}\rangle\right)
\end{eqnarray}
where the $\pm$ sign stands for $D'=D_1$ or $D^*$, respectively (We take the conventions $\mathcal{C} | D \rangle = + |\bar{D}\rangle$, $\mathcal{C} | D^* \rangle = - |\bar{D}^* \rangle$, $\mathcal{C} | D_1 \rangle = + |\bar{D}_1 \rangle$, where $\mathcal{C}$ is the charge conjugate transformation). For the $D^*\bar{D}^*$ decay, the $\mathcal{C}$-parity $C=+$ requires $L+S=\text{even}$, such that the flavor structure of the state is 
\begin{equation}\label{supeq:DstarDstar-opt}
    |D^*\bar{D}^*\rangle_{(C=+)}^{(I=0)}=\frac{1}{\sqrt{2}} \left(|D^{*+}D^{*-}\rangle+|D^{*0}\bar{D}^{*0}\rangle\right)_{(L=1)}^{(S=1)}
\end{equation}
with the relative momentum $L=1$ and the total spin $S=1$.

These flavor structures are taken into account both in the construction the two-meson operators and in the calculation of correlation functions. Now we consider the concrete spatial structure of the required two-meson operators. Since the quantum numbers $J,L,S,P$ are perfectly known for each decay mode, we name the two-meson operator by $\mathcal{O}_{AB}^M$ and omit the $JLSP$ subscripts in the following discussions and expressions. On the other hand, it is known that the $1^-$ ( $T_1^-$ ) operator has three components labeled by $i=1,2,3$ (corresponding to $x,y,z$ components, respectively). In practice, we use the third component ($i=3$) which corresponds to the $M=0$ case in Eq.~(\ref{eq:two-meson-sm}).  

The operators for the $AP$ mode ($D_1\bar{D}$ and $\chi_{c1}\eta_{(2)}$) are very simple. Here $\eta_{(2)}$ refers to the isoscalar pseudoscalar meson in the $N_f=2$ lattice QCD. Since the two mesons are in $S$-wave and both the $D_1\bar{D}$ and $\chi_{c1}\eta_{(2)}$ thresholds are very close to  $m_{\eta_{c1}}$, so we choose the momentum mode $\hat{k}=(0,0,0))$. On the other hand, we abbreviate the single meson operators by $A$ and $P$, respectively, in the explicit expressions of two-meson operators. This convention applies also for other decay modes. For the quantum numbers $(T_1^{+}, J=1, L=0, S=1, \hat{k}=(0,0,0))$, the operator is 
\begin{equation}\label{eq:AP-sm}
    \begin{aligned}
    \mathcal{O}_{AP}^{3}(0) = &  A^3(\vec{0}) P(\vec{0}).\\
    \end{aligned}
\end{equation}
Here we omit the constant factor $1/\sqrt{4\pi}$ that comes from $Y_{00}$. Actually for the momentum modes and the spin configurations involved in this work, the Clebsch-Gordan coefficients and the spherical harmonic functions result in the relative signs between different terms of a two-meson operator $\mathcal{O}_AB^3(\hat{k})$ apart from an overall constant factor. Since this constant factor can be cancelled out by taking a proper ratio of correlation functions and therefore is irrelevant to the physical results, we omit it throughout the construction of two-meson operators. 

The $VP$ mode is in $P$-wave and has only one particle composition, namely, $D^*\bar{D}$. The two mesons in $P$-wave have nonzero relative momentum. 
Since the momentum modes involved in this study are $\hat{k}=(n,0,0), (n,n,0), (n,n,n)$ types, the different orientations of the relative momentum 
are reflected by the signs of the its nonzero components. So we introduce three subscripts, which are different combinations of $+,-,0$, to the single meson operators. For example, $V^3_{+-0}$ means the third component of the operator for a vector meson of momentum $\vec{k}=\frac{2\pi}{La_s}(n,-n,0)$. Thus for the momentum mode $\hat{k}=(1,0,0)$, the $\mathcal{O}_{VP}$ operator of quantum number $(T_1^{-}, J=1, L=1, S=1)$ has four terms,
\begin{equation}\label{eq:PV100-sm}
    \mathcal{O}_{VP}^{3} = + V^1_{0+0} P_{0-0} - V^1_{0-0} P_{0+0} - V^2_{+00} P_{-00} + V^2_{-00} P_{+00}.
\end{equation}
In practice, we choose the momentum mode $\hat{k}=(1,1,1)$ on the L16 ensemble and $\hat{k}=(2,2,0)$ on the L24 ensemble, which make the energies of $D^*\bar{D}$ very close to $m_{\eta_{c1}}$. The explicit expression of $\mathcal{O}_{VP}^3$ of quantum numbers $(T_1^{-}, J=1, L=1, S=1$ for $\hat{k}=(2,2,0)$ is
\begin{widetext}
\begin{equation}\label{eq:VP022-sm}
    \begin{aligned}
    \mathcal{O}_{VP}^3 =& + V^1_{++0}P_{--0} + V^1_{+-0}P_{-+0} + V^1_{0++}P_{0--} + V^1_{0+-}P_{0-+} \\
                        & - V^1_{--0}P_{++0} - V^1_{-+0}P_{+-0} - V^1_{0--}P_{0++} - V^1_{0-+}P_{0+-} \\
                        & + V^2_{--0}P_{++0} + V^2_{-0-}P_{+0+} + V^2_{-0+}P_{+0-} + V^2_{-+0}P_{+-0} \\
                        & - V^2_{++0}P_{--0} - V^2_{+0+}P_{-0-} - V^2_{+0-}P_{-0+} - V^2_{+-0}P_{-+0} . 
    \end{aligned}
\end{equation}

For $\hat{k}=(1,1,1)$, one has 
\begin{equation}\label{eq:PV111-sm}
    \begin{aligned}
    \mathcal{O}_{VP}^3 =& + V^1_{+++}P_{---} + V^1_{++-}P_{--+} + V^1_{-+-}P_{+-+} + V^1_{-++}P_{+--}\\
                        & - V^1_{---}P_{+++} - V^1_{--+}P_{++-} - V^1_{+-+}P_{-+-} - V^1_{+--}P_{-++}\\
                        & + V^2_{---}P_{+++} + V^2_{--+}P_{++-} + V^2_{-+-}P_{+-+} + V^2_{-++}P_{+--}\\
                        & - V^2_{+++}P_{---} - V^2_{++-}P_{--+} - V^2_{+-+}P_{-+-} - V^2_{+--}P_{-++}.
    \end{aligned}
\end{equation}

Similarly, $PP$ operators are built for $\hat{k}=(n,0,0)$, $(n,n,0)$, and $(1,1,1)$ modes, respectively. The explicit expressions are 
\begin{equation}\label{eq:PP100-sm}
    \begin{aligned}
    \mathcal{O}_{PP}^3 = &+ P_{00+}P'_{00-} - P_{00-}P'_{00+}
\end{aligned}
\end{equation}
for $\hat{k}=(n,0,0)$, 
\begin{equation}\label{eq:PP111-sm}
    \begin{aligned}
    \mathcal{O}_{PP}^3 =&   + P_{+++}  P'_{---}
                            - P_{++-}  P'_{--+}
                            + P_{+-+}  P'_{-+-}
                            - P_{+--}  P'_{-++}\\
                           &+ P_{-++}  P'_{+--}
                            - P_{-+-}  P'_{+-+}
                            + P_{--+}  P'_{++-}
                            - P_{---}  P'_{+++} 
    \end{aligned}
\end{equation}
for $\hat{k}=(1,1,1)$, and 
\begin{equation}\label{eq:PP022-sm}
    \begin{aligned}
    \mathcal{O}_{PP}^3 =&
        - P_{-0-}P'_{+0+}
        + P_{-0+}P'_{+0-} 
        - P_{0--}P'_{0++}
        + P_{0-+}P'_{0+-} \\
       &- P_{0+-}P'_{0-+}
        + P_{0++}P'_{0--} 
        - P_{+0-}P'_{-0+}
        + P_{+0+}P'_{-0-}, 
    \end{aligned}
\end{equation}
for $\hat{k}=(n,n,0)$.
\begin{figure}
    \centering
    \includegraphics[width=0.45\linewidth]{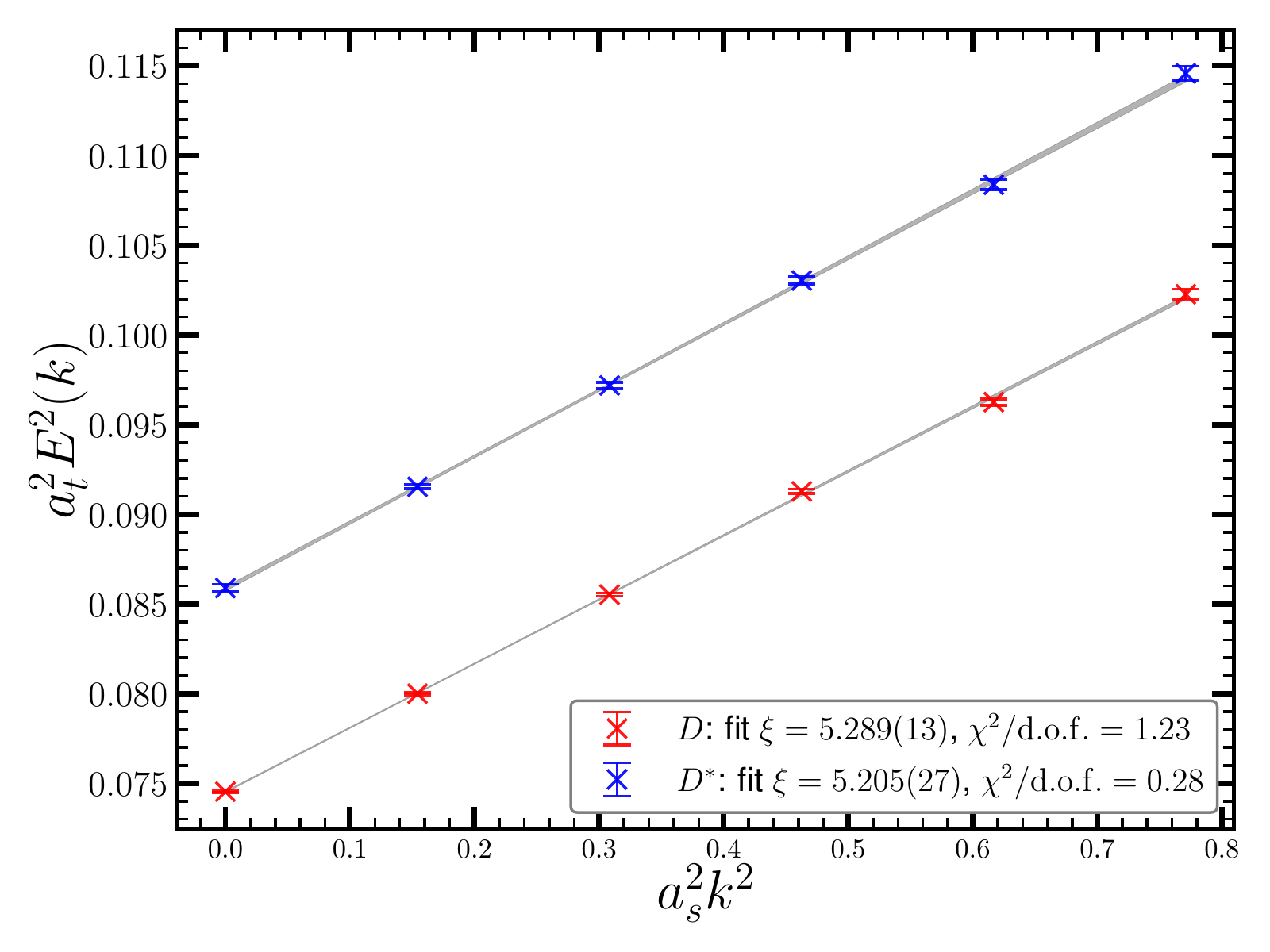}
    \includegraphics[width=0.45\linewidth]{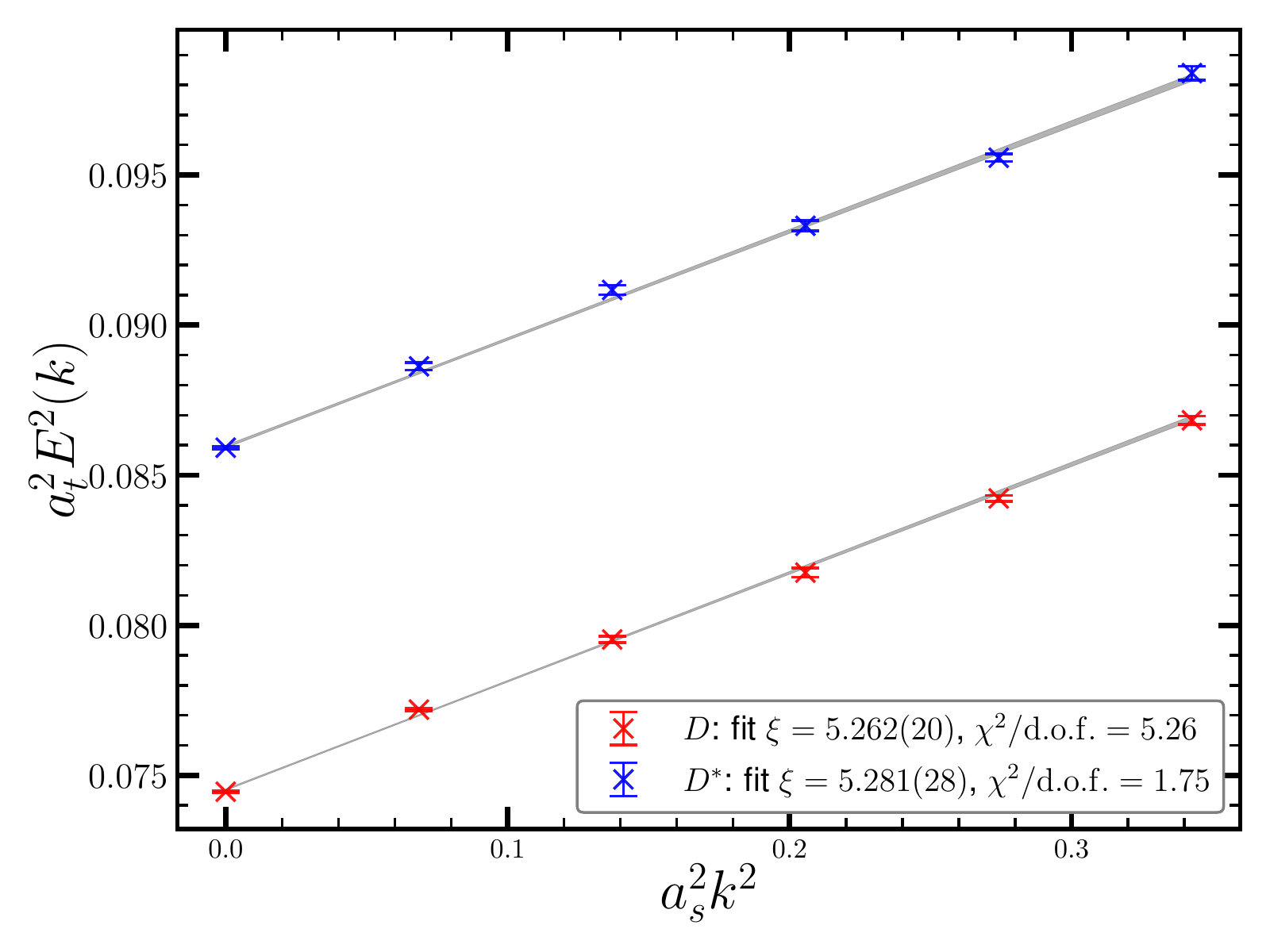}
    \caption{Dispersion relation of $D$ and $D^*$ mesons on gauge ensembles L16 and L24 respectively. The derived aspect ratios are compatible with the tuned value $\xi\approx 5.3$.}
    \label{fig:reply-DDs-disp}
\end{figure}

In the $VV$ mode we only consider the open-charm $D^*\bar{D}^*$ decay since the $J/\psi\omega(\phi)$ decay is strongly suppressed by the OZI rule. For this particle-antiparticle pair, the $\mathcal{C}$-parity $C=+$ requires $L+S=\text{even}$, so we construct the $D^*\bar{D}^*$ operators of quantum numbers $(T_1^{-}, J=1, L=1, S=1)$. Let $V^i$ and $\bar{V}^i$ be the operators for $D^*$ and $\bar{D}^*$, respectively, then for the momentum mode $\hat{k}=(n,0,0)$, we have 
\begin{equation}\label{eq:VV100-sm}
    \begin{aligned}
      \mathcal{O}_{VV}^3 =& - V^1_{+00}\bar{V}^3_{-00} + V^1_{-00}\bar{V}^3_{+00} 
                            - V^2_{0+0}\bar{V}^3_{0-0} + V^2_{0-0}\bar{V}^3_{0+0}\\ 
                          & + V^3_{+00}\bar{V}^1_{-00} - V^3_{-00}\bar{V}^1_{+00} 
                            + V^3_{0+0}\bar{V}^2_{0-0} - V^3_{0-0}\bar{V}^2_{0+0}.
    \end{aligned}
\end{equation}

For the momentum mode $\hat{k}=(n,n,0)$, the $VV$ operator reads,
\begin{equation}\label{eq:VVnn0-sm}
    \begin{aligned}
    \mathcal{O}_{VV}^3 =
    &   - V^{1}_{++0}     \bar{V}^{3}_{--0}
        - V^{1}_{+0+}     \bar{V}^{3}_{-0-}
        - V^{1}_{+0-}     \bar{V}^{3}_{-0+}
        - V^{1}_{+-0}     \bar{V}^{3}_{-+0} \\
    &   + V^{1}_{-+0}     \bar{V}^{3}_{+-0}
        + V^{1}_{-0+}     \bar{V}^{3}_{+0-}
        + V^{1}_{-0-}     \bar{V}^{3}_{+0+}
        + V^{1}_{--0}     \bar{V}^{3}_{++0} \\
    &   - V^{2}_{++0}     \bar{V}^{3}_{--0}
        + V^{2}_{+-0}     \bar{V}^{3}_{-+0}
        - V^{2}_{0++}     \bar{V}^{3}_{0--}
        - V^{2}_{0+-}     \bar{V}^{3}_{0-+} \\
    &   + V^{2}_{0-+}     \bar{V}^{3}_{0+-}
        + V^{2}_{0--}     \bar{V}^{3}_{0++}
        - V^{2}_{-+0}     \bar{V}^{3}_{+-0}
        + V^{2}_{--0}     \bar{V}^{3}_{++0} \\
    &   + V^{3}_{++0}     \bar{V}^{2}_{--0}
        + V^{3}_{++0}     \bar{V}^{1}_{--0}
        + V^{3}_{+0+}     \bar{V}^{1}_{-0-}
        + V^{3}_{+0-}     \bar{V}^{1}_{-0+} \\
    &   - V^{3}_{+-0}     \bar{V}^{2}_{-+0}
        + V^{3}_{+-0}     \bar{V}^{1}_{-+0}
        + V^{3}_{0++}     \bar{V}^{2}_{0--}
        + V^{3}_{0+-}     \bar{V}^{2}_{0-+} \\
    &   - V^{3}_{0-+}     \bar{V}^{2}_{0+-}
        - V^{3}_{0--}     \bar{V}^{2}_{0++}
        + V^{3}_{-+0}     \bar{V}^{2}_{+-0}
        - V^{3}_{-+0}     \bar{V}^{1}_{+-0} \\
    &   - V^{3}_{-0+}     \bar{V}^{1}_{+0-}
        - V^{3}_{-0-}     \bar{V}^{1}_{+0+}
        - V^{3}_{--0}     \bar{V}^{2}_{++0}
        - V^{3}_{--0}     \bar{V}^{1}_{++0}.
    \end{aligned}
\end{equation}

The $VV$ operator for the momentum mode $\hat{k}=(1,1,1)$ is expressed as
\begin{equation}\label{eq:VV111-sm}
    \begin{aligned}
    \mathcal{O}_{VV} = 
                    -& V^{1}_{+++}\bar{V}^{3}_{---} - V^{1}_{++-}\bar{V}^{3}_{--+} 
                    - V^{1}_{+-+}\bar{V}^{3}_{-+-} - V^{1}_{+--}\bar{V}^{3}_{-++} \\
                    +& V^{1}_{-++}\bar{V}^{3}_{+--} + V^{1}_{-+-}\bar{V}^{3}_{+-+} 
                    + V^{1}_{--+}\bar{V}^{3}_{++-} + V^{1}_{---}\bar{V}^{3}_{+++}\\
                    -& V^{2}_{+++}\bar{V}^{3}_{---} - V^{2}_{++-}\bar{V}^{3}_{--+} 
                    + V^{2}_{+-+}\bar{V}^{3}_{-+-} + V^{2}_{+--}\bar{V}^{3}_{-++} \\
                    -& V^{2}_{-++}\bar{V}^{3}_{+--} - V^{2}_{-+-}\bar{V}^{3}_{+-+} 
                    + V^{2}_{--+}\bar{V}^{3}_{++-} + V^{2}_{---}\bar{V}^{3}_{+++} \\
                    +& V^{3}_{+++}\bar{V}^{2}_{---} + V^{3}_{+++}\bar{V}^{1}_{---} 
                    + V^{3}_{++-}\bar{V}^{2}_{--+} + V^{3}_{++-}\bar{V}^{1}_{--+} \\
                    -& V^{3}_{+-+}\bar{V}^{2}_{-+-} + V^{3}_{+-+}\bar{V}^{1}_{-+-} 
                    - V^{3}_{+--}\bar{V}^{2}_{-++} + V^{3}_{+--}\bar{V}^{1}_{-++} \\
                    +& V^{3}_{-++}\bar{V}^{2}_{+--} - V^{3}_{-++}\bar{V}^{1}_{+--} 
                    + V^{3}_{-+-}\bar{V}^{2}_{+-+} - V^{3}_{-+-}\bar{V}^{1}_{+-+} \\
                    -& V^{3}_{--+}\bar{V}^{2}_{++-} - V^{3}_{--+}\bar{V}^{1}_{++-} 
                    - V^{3}_{---}\bar{V}^{2}_{+++} - V^{3}_{---}\bar{V}^{1}_{+++}.
    \end{aligned}
\end{equation}

Since the $D^{(*)}\bar{D}^{(*)}$ decay modes involve the $D^{(*)}$ and $\bar{D}^{(*)}$ mesons in flight, the dispersion relations of $D$ and $D^*$ are checked on the two ensembles L16 and L24 through the relation
\begin{equation}
    E^2(k) a_t^2=m^2 a_t^2 + \frac{1}{\xi^2} k^2 a_s^2.
\end{equation}
As shown in Fig.~\ref{fig:reply-DDs-disp}, this relation is statisfied with the values of $\xi$ being compatible with the tuned value $\xi\approx 5.3$.
\end{widetext}

\subsection{The Parametrization of correlation functions}\label{sec:appendix:parametrization-correlation}
As for the decay process $\eta_{c1}\to AB$, we are interested in the following correlation function
\begin{equation}
    C^{ij}_{AB,\eta_{c1}}(\hat{k},t)\equiv\langle 0|\mathcal{O}^i_{AB}(\hat{k},t)\mathcal{O}^{j,\dagger}_{\eta_{c1}}(0)|0\rangle,
\end{equation}
where $\mathcal{O}^i_{\eta_{c1}}$ is the interpolating operator that annihilates a $\eta_{c1}$ state, $\mathcal{O}^i_{AB}(\hat{k})(t)$ is any one of  the two-particle operators defined in Eqs.~(\ref{eq:AP-sm}, \ref{eq:PV100-sm}, \ref{eq:PV111-sm}, \ref{eq:VVnn0-sm}, \ref{eq:VV111-sm}, \ref{eq:PP100-sm}, \ref{eq:PP111-sm},
\ref{eq:PP022-sm}) that annihilates a $AB$ state with the same quantum number as that of $\eta_{c1}$. Actually, each term $\mathcal{O}_{AB}^i(\vec{k})$ of the operator $\mathcal{O}^i_{AB}(\hat{k})$ with a specifically oriented momentum $\vec{k}$ and a specific index combination of $\mathcal{O}_A$ and $\mathcal{O}_B$ (if available), 
\begin{equation}\label{eq:oper-AB-sm}
    \mathcal{O}_{AB}^i(\vec{k}) = \left[\mathcal{O}_A(\vec{k})\circ \mathcal{O}_B(-\vec{k})\right]^i, 
\end{equation}
contributes uniformly to $C^{ij}_{AB,\eta_{c1}}(\hat{k},t)$ by a quantity
\begin{equation}\label{eq:corr-veck-sm}
    C^{ij}_{AB,\eta_{c1}}(\vec{k},t)\equiv\langle 0|\mathcal{O}^i_{AB}(\vec{k},t)\mathcal{O}^{j,\dagger}_{\eta_{c1}}(0)|0\rangle,
\end{equation}
with $\vec{k}$ being the spatial momentum of $A$ in the rest frame of $\eta_{c1}$. In Eq.~(\ref{eq:oper-AB-sm}) the symbol `$\circ$' stands for the specific combination in the explicit expression of $\mathcal{O}^i_{AB}(\hat{k})$ that results from Eq.~(\ref{eq:two-meson-sm}). 

Let $\ket{A(\vec{k})B(-\vec{k})}$ be the state generated by $\mathcal{O}^{i,\dagger}_{AB}(\hat{k})$, then in the space of $\ket{\eta_{c1}}$ and $\ket{A(\vec{k})B(-\vec{k})}$,  
the temporal transfer matrix is expressed as 
\begin{equation}\label{eq:transfer-sm}
    \hat{T}=e^{-a_t\hat{H}}=e^{-a_t\bar{E}} \left(
    \begin{array}{cc}
    e^{-a_t\Delta/2} & a_t x_{AB}\\
    a_t x_{AB} & e^{+a_t\Delta/2}
    \end{array}\right).
\end{equation}
where $\bar{E}=(m_{\eta_{c1}}+E_{AB})/2$ and $\Delta=m_{\eta_{c1}}-E_{AB}$ have been defined with $E_{AB}=E_A+E_B$. Note that this expression assumes the state normalizations $\langle \eta_{c1}|\eta_{c1}\rangle=1$ and $\langle AB|AB\rangle=1$. Without losing the generality, in the following discussion we use the state normalization conditions 
\begin{eqnarray}
    \mathcal{N}_{\eta_{c1}} &\equiv& \langle \eta_{c1},\lambda|\eta_{c1},\lambda\rangle \nonumber\\
    \mathcal{N}_{AB}&\equiv&\langle AB;(\lambda'\lambda''),\vec{k}|AB;(\lambda'\lambda''),\vec{k}\rangle,
\end{eqnarray}
where $\lambda$ is the polarization index of $\eta_{c1}$, while $\lambda'$ and $\lambda''$ represent the the polarization indices (if available) for the final state mesons $A$ and $B$, respectively. If we assume $\mathcal{O}_{\eta_{c1}}^i$ does not couple to $|AB\rangle$ states and $\mathcal{O}_{AB}^i$ does not couple to the $|\eta_{c1}\rangle$ state, then the $\eta_{c1}-AB$ transition takes place at any time slice $t'\in[0,t]$, such that 
\begin{widetext}
\begin{eqnarray}\label{eq:corr-HAB-sm}
C^{ij}_{AB,\eta_{c1}}(t)&\sim&\sum\limits_{\lambda (\lambda'\lambda'')}\frac{1}{\mathcal{N}_{\eta_{c1}}\mathcal{N}_{AB}} Z_{AB,i}^{(\lambda'\lambda'')} Z_{\eta_{c1,j}}^{\lambda,*} \langle AB;(\lambda'\lambda''),\vec{k}|e^{-a_t\hat{H}t}|\eta_{c1},\lambda \rangle\nonumber\\
&\sim&\sum\limits_{\lambda(\lambda'\lambda'')} a_t x_{AB}^{(\lambda'\lambda'')\lambda}\frac{1}{\mathcal{N}_{\eta_{c1}}\mathcal{N}_{AB}}Z_{AB,i}^{(\lambda'\lambda'')}Z_{\eta_{c1},j}^{\lambda,*}(\lambda)\nonumber\sum\limits_{t_1=0}^{t} e^{-a_t E_{AB}(t-t_1-1)} e^{-a_tm_{\eta_{c1}}t_1}\nonumber\\
&\sim& -\frac{\sinh (a_t\Delta t/2)}{\sinh (a_t\Delta/2)} e^{-a_t\bar{E} t}\sum\limits_{\lambda(\lambda'\lambda'')} a_t x_{AB}^{(\lambda'\lambda'')\lambda} \frac{1}{\mathcal{N}_{\eta_{c1}}\mathcal{N}_{AB}}Z_{AB,i}^{(\lambda'\lambda'')}Z_{\eta_{c1},j}^{\lambda,*}\nonumber\\
&\to& -t\left(1+\frac{1}{24}(a_t\Delta t)^2\right) e^{-a_t\bar{E} t}\sum\limits_{\lambda(\lambda'\lambda'')} a_t x_{AB}^{(\lambda'\lambda'')\lambda} \frac{1}{\mathcal{N}_{\eta_{c1}}\mathcal{N}_{AB}}Z_{AB,i}^{(\lambda'\lambda'')}Z_{\eta_{c1,j}}^{\lambda,*},
\end{eqnarray}
\end{widetext}
where $x_{AB}$ is now replaced by the transition amplitude
\begin{equation}
    x_{AB}^{(\lambda'\lambda'')\lambda}(\vec{k})\approx \langle AB;(\lambda'\lambda''),\vec{k}|\hat{H}|\eta_{c1}, \lambda \rangle,
\end{equation}
with proper state normalizations, $Z_{\eta_{c1,j}}^{\lambda}$ and $Z_{AB,i}^{(\lambda'\lambda'')}$ are the overlapping matrix elements
\begin{equation}
\begin{aligned}
    Z_{\eta_{c1},i}^{\lambda}=&\langle 0|\mathcal{O}_{\eta_{c1}}^i|\eta_{c1},\lambda \rangle
    \equiv Z_{\eta_{c1}}\epsilon^i_\lambda(\vec{0})\\
    Z_{AB,i}^{(\lambda'\lambda'')}=&\langle 0|\mathcal{O}^i_{AB}|AB;(\lambda'\lambda''),\vec{k}\rangle.\\
\end{aligned}
\end{equation}
The parametrization of $Z_{AB,i}^{(\lambda'\lambda'')}$ is not so straightforward as that of $Z_{\eta_{c1,i}}^{\lambda}$. To the first order of the Bonn approximation, the two-particle state can be expressed as the direct product of the state vector of $A$ and $B$, namely, 
\begin{equation}
    |AB;(\lambda',\lambda''),\vec{k}\rangle\approx |A,(\lambda'),\vec{k}\rangle \otimes |B,(\lambda'),-\vec{k}\rangle,
\end{equation}
such that $\mathcal{N}_{AB}$ and $Z_{AB,i}^{(\lambda'\lambda'')}$ are approximated by 
\begin{eqnarray}
    \mathcal{N}_{AB}&\approx& \mathcal{N}_A \mathcal{N}_B\nonumber\\
    Z_{AB,i}^{(\lambda'\lambda'')}&\approx& Z_{A}Z_{B} \epsilon^i_{(\lambda'\lambda'')}(\vec{k},AB),
\end{eqnarray}
where $Z_A$ and $Z_B$ are defined through 
\begin{equation}
    \langle 0|\mathcal{O}_X^{(i)}(\vec{k})|X;\lambda',\vec{k}\rangle=Z_X \epsilon_{(\lambda')}^{(i)}(\vec{k})
\end{equation}
with $X$ referring to $A$ or $B$, and $\epsilon^i_{(\lambda'\lambda'')}(\vec{k},AB)$ is the polarization of $AB$ defined by 
\begin{equation}
    \epsilon^i_{(\lambda'\lambda'')}(\vec{k},AB)=\left[\vec{\epsilon}_{(\lambda')}(\vec{k})\circ\vec{\epsilon}_{(\lambda'')}(-\vec{k})\right]^i.
\end{equation}
The meaning of the symbol ``$\circ$" here is the same as that in Eq.~(\ref{eq:oper-AB-sm}). The brackets of the superscripts and subscripts mean
that they are void for pseudoscalar mesons and the normal meanings for (axial) vector mesons. 
The factors $Z_{\eta_{c1}}$, $Z_A$ and $Z_B$ are also encoded in the individual correlation functions of operators $\mathcal{O}_{\eta_{c1}}$, $\mathcal{O}_{A}^{(i)}(\vec{k})$ and $\mathcal{O}_B^{(i)}(\vec{k})$ (no superscripts for pseudoscalar mesons),
\begin{equation}
    C_{XX}^{(ii)}(t;\vec{k})=\langle 0|\mathcal{O}_X^{(i)}(\vec{k},t)\mathcal{O}_X^{(i)\dagger}(\vec{k},0)|\rangle 
    \to  \frac{Z_X^2}{\mathcal{N}_X} \mathcal{P}_{ii}^X(\vec{k}) e^{-a_t E_X t},
\end{equation}
where the normalization condition $\mathcal{N}_X=2E_X L^3$ is used for single particle states $X=\eta_{c1},A,B$, and $\mathcal{P}_X^{(ij)}$ is unity for pseudoscalars and 
\begin{equation}
    \mathcal{P}_X^{ij}(\vec{k})=\sum\limits_\lambda \epsilon^i_\lambda(\vec{k})\epsilon_\lambda^{j*}(\vec{k})=\delta_{ij}+\frac{k_i k_j}{m_X^2},
\end{equation}
comes from the completeness of polarization vectors of (axial) vector mesons. Consequently, we have the ratio function
\begin{widetext}
\begin{eqnarray}\label{eq:ratio-sm}
    R_{AB}(\vec{k},t)&=&\frac{C^{33}_{AB,\eta_{c1}}(\vec{k},t)}{\sqrt{[C_{AA}(\vec{k},t)C_{BB}(-\vec{k},t)]^{33} C^{33}_{\eta_{c1},\eta_{c1}}(\vec{0},t)}}\nonumber\\
        &\approx& t\left(1+\frac{1}{24}(a_t\Delta)^2 t^2\right)\frac{1}{\sqrt{8L^3 m_{\eta_{c1}}E_A E_B}}\nonumber\\
        &\times&\sum\limits_{\lambda(\lambda'\lambda'')} \frac{a_t x^{(\lambda'\lambda'')\lambda}_{AB}[\epsilon_{(\lambda'\lambda'')}^3(\vec{k},AB)\epsilon^{3}_\lambda(\vec{0})]}{\sqrt{\mathcal{P}_{A}(\vec{k}) \mathcal{P}_{B}(-\vec{k}) \mathcal{P}_{\eta_{c1}}^{33}(\vec{0})}}~~~(t\gg 1~~\text{and}~~(a_t\Delta) t\ll 1),
\end{eqnarray}
\end{widetext}
which is free of the factors $Z_{\eta_{c1}}$, $Z_A$ and $Z_B$. 
 It is easily seen that the transition amplitude
\begin{equation}
    x_{AB}^{(\lambda'\lambda'')\lambda}\approx -\langle AB;\lambda'\lambda'',\vec{k}|\hat{H}|\eta_{c1}, \lambda \rangle
\end{equation}
can be derived from the slope of $R_{AB}(\vec{k},t)$ with respect to $t$. 

In practice, we average contribution of Eq.~(\ref{eq:corr-veck-sm}) over 
all the terms of $\mathcal{O}_{AB}^i(\hat{k})$ to increase the statistics. The correlation functions of $C_{AA}(\vec{k},t)$ and $C_{BB}(-\vec{k},t)$ are also averaged accordingly.

\subsection{Amplitudes}
The two-body decay $\eta_{c1}\to AB$ includes closed-charm decays $\chi_{c1}\eta(\eta')$, $\eta_c \eta(\eta')$, $J/\psi\omega(\phi)$, and open-charm modes $D_{(s)}^*\bar{D}_{(s)}$, $D_{(s)}^*\bar{D}_{(s)}^*$, $D\bar{D}_1(2420)$, etc. The effective Lagrangian for closed-charm decays can be written as
\begin{widetext}
\begin{equation}\label{eq:L1-sm}
        \mathcal{L}_\mathrm{I}^\mathrm{cc}\sim -g_{\chi\eta}m_{\eta_{c1}} H_\mu A^\mu \eta-ig_{\eta_c\eta} H_\mu \eta_c \overleftrightarrow{\partial}^\mu \eta
        +i H_\mu \left(g\psi_\nu\partial^\nu \omega^{\mu}+g'\omega_\nu \partial^\nu \psi^\mu+g_0\psi_\nu \overleftrightarrow{\partial}^\mu \omega^{\nu}\right),
\end{equation}
where $\overleftrightarrow{\partial}$ represents $\overleftarrow{\partial}-\overrightarrow{\partial}$, and the fields $H_\mu$, $\chi_\mu$, $\eta_c$, $\eta$, $\psi_\mu$ and $\omega_\mu$ are for $\eta_{c1}$, $\chi_{c1}$, $\eta_c$, $\eta(\eta')$, $J/\psi$ and $\omega$ mesons, respectively. When the isospin doublet fields $D_{1\mu}$, $D_\mu$ and $D$ are introduced for $D_1$, $D^*$ and $D$ mesons of the flavor wave functions $(|c\bar{d}\rangle,-|c\bar{u}\rangle)^T$, the effective Lagrangian for open-charm decays reads
\begin{equation}\label{eq:L2-sm}
\begin{aligned}
     \mathcal{L}_\mathrm{I}^\mathrm{oc}&\sim  g_{D_1D} m_{\eta_{c1}} H_\mu\frac{1}{2}\left(D_1^{\mu,\dagger} D+D^\dagger D_1^\mu\right)
     +g_{D^*\bar{D}^*} H^\mu \frac{i}{\sqrt{2}}\left(D^{\nu,\dagger}\partial_\nu D_\mu+\partial_\nu D^{\dagger}_\mu D^\nu\right)\\
     &+\frac{g_{D^*\bar{D}}}{m_{\eta_{c1}}} \epsilon^{\mu\nu\rho\sigma} (\partial_\mu H_\nu) \frac{1}{2}
     \left[(\partial_\rho D_{\sigma}^{\dagger}) D- D^\dagger(\partial_\rho D_\sigma)\right].
\end{aligned}
\end{equation}
\end{widetext}
According to the effective Lagrangian, the tree-level decay amplitudes read
\begin{equation}\label{eq:amplitude-sm}
    \begin{aligned}
        x_{AP}^{\lambda'\lambda}=& {g}_{AP} m_{\eta_{c1}} \vec{\epsilon}_\lambda(\vec{0})\cdot \vec{{\epsilon}}^{~*}_{\lambda'}(\vec{k}),\\
        x_{PP}^\lambda=& 2{g}_{PP} \vec{{\epsilon}}_\lambda(\vec{0})\cdot \vec{k},\\
        x_{VP}^{\lambda'\lambda}=& {g}_{VP} \vec{\epsilon}_\lambda(\vec{0})\cdot(\vec{\epsilon}_{\lambda'}^{~*}(\vec{k})\times \vec{k}), \\
        x_{D^*\bar{D}^*}^{\lambda'\lambda''\lambda}=& 2{g}_{D^*\bar{D}^*}\vec{\epsilon}_\lambda(\vec{0})\cdot\left(\vec{k}\times \left[\vec{\epsilon}^{~*}_{\lambda'}(\vec{k})\times \vec{\epsilon}^{~*}_{\lambda''}(-\vec{k})\right]\right),\\
    \end{aligned}
\end{equation}
where $\vec{\epsilon}_\lambda(\vec{0})$, $\vec{\epsilon}_{\lambda'}(\vec{k})$ and $\vec{\epsilon}_{\lambda''}(-\vec{k})$ are the spatial components of the polarization vectors of the initial state $\eta_{c1}$ and the vector mesons in the final state, respectively. As shown in Eq.~(\ref{eq:amplitude-sm}), the 
amplitude $x_{AB}$ for the decay process $\eta_{c1}\to AB$ is encoded in the ratio function $R_{AB}(\vec{k},t)$ and can be extracted from its slope in the time range where it behaves as a linear function. In the practical data analysis procedure, $R_{AB}(\vec{k},t)$ is fitted by the function form
\begin{equation}\label{eq:ratio-model-sm}
    R_{AB}(\vec{k},t)=r_0+r_1 t+r_3 t^3
\end{equation}
and the fitted value of the parameter $r_1$ is proportional to the amplitude $x_{AB}$. Before giving the numerical details of the extraction of $r_1$, we derive the explicit relation between $r_1$ and the effective coupling $g_{AB}$ mode by mode using Eq.~(\ref{eq:amplitude-sm})as follows. 
\begin{widetext}
\begin{itemize}
    \item \textbf{$AP$ mode}: On our lattices, the mass thresholds of $\chi_{c1}\eta_{(2)}$ and $D_1\bar{D}$ are very close to $m_{\eta_{c1}}$, and the final $AP$ states are in $S$-wave ($L=0$), so we set $\vec{k}=\vec{0}$. According to the operator structure in Eq.~(\ref{eq:AP-sm}), one has 
   \begin{eqnarray}
     r_1 &=& \frac{1}{\sqrt{8L^3 m_{\eta_{c1}}E_A E_P}} \sum\limits_{\lambda\lambda'} a_t x_{PA}^{\lambda'\lambda} \frac{\epsilon^3_{\lambda'}(\vec{k})\epsilon^{3*}_\lambda(\vec{0})}{\sqrt{\mathcal{P}_{ii}^{A}(\vec{k})\mathcal{P}^{\eta_{c1}}_{jj}(\vec{0})}}\nonumber\\
     &=&   \frac{{g}_{AP} m_{\eta_{c1}} a_t}{\sqrt{8L^3 m_{\eta_{c1}}E_A E_P}} \sum\limits_{\lambda\lambda'}\frac{\epsilon^3_{\lambda'}(\vec{k})\epsilon^{3*}_\lambda(\vec{0}) \epsilon^{\mu}_\lambda(\vec{0}) \epsilon^*_{\mu,\lambda'}(\vec{k})}{\sqrt{\mathcal{P}_{ii}^{A}(\vec{k})\mathcal{P}^{\eta_{c1}}_{ii}(\vec{0})}},\nonumber\\
     &=&   \frac{{g}_{AP} m_{\eta_{c1}} a_t}{\sqrt{8L^3 m_{\eta_{c1}} E_A E_P}} \frac{\delta^{\mu}_{3}(\delta_{\mu 3}+\frac{k_\mu k_3}{m_A^2})}{\sqrt{(1+\frac{k_3^2}{m_A^2})\cdot 1}} \nonumber\\
     &=&   \frac{{g}_{AP} m_{\eta_{c1}} a_t}{\sqrt{8L^3 m_{\eta_{c1}}E_A E_P}}.
\end{eqnarray} 
Or equivalently we have
\begin{equation}
    {g}_{AP} = \frac{\sqrt{8L^3 m_{\eta_{c1}}E_A E_B}}{m_{\eta_{c1}} a_t} r_1.
\end{equation}

    \item \textbf{$PP$ mode}: Similarly to the case of $AP$ mode, for $\vec{k}=\frac{2\pi}{La_s}(0,0,1)$ (the first term of Eq.~(\ref{eq:PP100-sm})), we have 
\begin{equation}
      r_1=  2 a_t {g}_{PP} \frac{1}{\sqrt{8L^3 m_{\eta_{c1}}E_P E_{P'}}} \sum\limits_{\lambda} \frac{   \epsilon^{i}_\lambda(\vec{0})  k_i \epsilon^{3*}_\lambda(\vec{0})}{\sqrt{\mathcal{P}^{\eta_{c1}}_{33}(\vec{0})}} 
       =   {g}_{PP} \frac{2 k_3 a_t}{\sqrt{8L^3 m_{\eta_{c1}}E_P E_{P'}}}, 
\end{equation}
where $k^3$ is the $z$-component of $\vec{k}$. Finally we have 
\begin{equation}
    {g}_{PP} =  \frac{\sqrt{8L^3 m_{\eta_{c1}}E_P E_{P'}}}{2 k_3 a_t} r_1
\end{equation}

 \item \textbf{$VP$ mode}: The first terms of Eq.~(\ref{eq:PV100-sm}), Eq.~(\ref{eq:VP022-sm}), and Eq.~(\ref{eq:PV111-sm}) give the same expression of $r_1$
\begin{equation}
    \begin{aligned}
      r_1= \frac{1}{\sqrt{8L^3 m_{\eta_{c1}}E_V E_P}} \sum\limits_{\lambda\lambda'} a_t x_{VP}^{\lambda'\lambda}\frac{\epsilon^1_{\lambda'}(\vec{k}) \epsilon^{3}_\lambda(\vec{0}) }{\sqrt{\mathcal{P}_{11}^{V}(\vec{k})}}
      = \frac{{g}_{VP}}{\sqrt{8L^3 m_{\eta_{c1}}E_V E_P}} \frac{k_2 a_t}{\sqrt{1+\frac{k_1 k_1}{m_V^2}}}, 
    \end{aligned}
\end{equation}
from which the effective coupling $g_{VP}$ is determined through
\begin{equation}
    {g}_{VP} = \sqrt{1+\frac{k_1 k_1}{m_V^2}}\frac{ {\sqrt{8L^3 m_{\eta_{c1}}E_V E_P}} } {k_2 a_t} r_1
\end{equation}

\item \textbf{$VV$ mode}: Here we only consider the open-charm mode $D^*\bar{D}^*$, which must be in the $L=1$ and $S=1$ state. The first terms of Eq.~(\ref{eq:VV100-sm}), Eq.~(\ref{eq:VV111-sm}) and Eq.~(\ref{eq:VVnn0-sm}) gives 
\begin{equation}
     r_1 = \frac{1}{\sqrt{8L^3 m_{\eta_{c1}}E_{D^*}^2}} \sum\limits_{\lambda\lambda'\lambda''} a_t x_{D^*\bar{D}^*}^{\lambda'\lambda''\lambda} \frac{\epsilon^1_{\lambda'}(\vec{k})\epsilon^3_{\lambda''}(-\vec{k})\epsilon^{3*}_\lambda(\vec{0})}{\sqrt{\mathcal{P}_{11}^{D^*}(\vec{k})\mathcal{P}^{\bar{D^*}}_{33} \mathcal{P}^{\eta_{c1}}_{33}(\vec{0})}}.
\end{equation}
Then for $\vec{k}=\frac{2\pi}{La_s}(n,n,n)$,  $r_1$ is expressed as  
\begin{equation}
    r_1=2 g_{D^*\bar{D}^*}  k_1 a_t \left(1+\frac{\vec{k}^2}{m_{D^*}^2}\right)\left(1+\frac{k^2_1}{m_{D^*}^2}\right)^{-1},
\end{equation}
For $\vec{k}=\frac{2\pi}{La_s}(n,0,0)$ and $\vec{k}=\frac{2\pi}{La_s}(n,n,0)$, one has 
\begin{equation}
    r_1 = 2g_{D^*\bar{D}^*}  k_1 a_t \left(1+\frac{\vec{k}^2}{m_{D^*}^2}\right)\left(1+\frac{k^2_1}{m_{D^*}^2}\right)^{-1/2}.
\end{equation}
Thus, we obtain $g_{D^*\bar{D}^*}$ at different momenta $\vec{k}$.
\end{itemize}

With the effective coupling $g_{AB}$ determined from $r_1$ for each decay mode, after averaging over the initial state polarization and the summation of the final state polarization, one can obtain the explicit expression of the decay amplitude squared
\begin{equation}
    \overline{|\mathcal{M}(\eta_{c1}\to AB)|^2}=\frac{1}{3}\sum_{\lambda(\lambda'\lambda'')} \left|x_{AB}^{(\lambda'\lambda'')\lambda}\right|^2
\end{equation}
which enters calculation of the partial decay width directly 
\begin{equation}
\Gamma_{AB}=\frac{1}{8\pi}\frac{k_\text{ex}}{m_{\eta_{c1}}^2}\overline{|\mathcal{M}(\eta_{c1}\to AB)|^2},     
\end{equation}
where $k_\text{ex}$ is the decay momentum determined by the experimental mass values of $A$ and $B$ along with a given $m_{\eta_{c1}}$, namely, 
\begin{equation}
    k_\text{ex}=\frac{1}{2m_{\eta_{c1}}}\lambda(m_{\eta_{c1}}^2, m_A^2, m_B^2)^{1/2}\equiv \frac{1}{2m_{\eta_{c1}}} \left(m_{\eta_{c1}}^4+m_A^4+m_B^4-2m_{\eta_{c1}}^2m_A^2-2m_{\eta_{c1}}^2m_B^2-2m_A^2m_B^2\right)^{1/2}.
\end{equation}
Specifically for the decay modes considered in this study, the explicit expressions of $\overline{|\mathcal{M}|^2}$ are
\begin{equation}
  \begin{aligned}
    \overline{| \mathcal{M}(\eta_{c1}\to AP)|^2} =& \frac{1}{3} {g}_{AP}^2 m_{\eta_{c1}}^2 (3+\frac{{k}_\text{ex}^2}{m_A^2}), \\
    \overline{| \mathcal{M}(\eta_{c1}\to PP)|^2} =&  \frac{4}{3} {g}_{PP}^2 {k}_\text{ex}^2, \\
    \overline{| \mathcal{M}(\eta_{c1}\to VP)|^2} =&  \frac{2}{3} {g}_{VP}^2 {k}_\text{ex}^2,\\
    \overline{| \mathcal{M}(\eta_{c1}\to D^*\bar{D}^*)|^2} =& \frac{4}{3} g^2 {k}_\text{ex}^2 \frac{m_{\eta_{c1}}^2}{m_{D^*}^2}.  \\
  \end{aligned}
\end{equation}
\end{widetext}

\begin{figure}[!ht]
    \centering
    \includegraphics[width=1\linewidth]{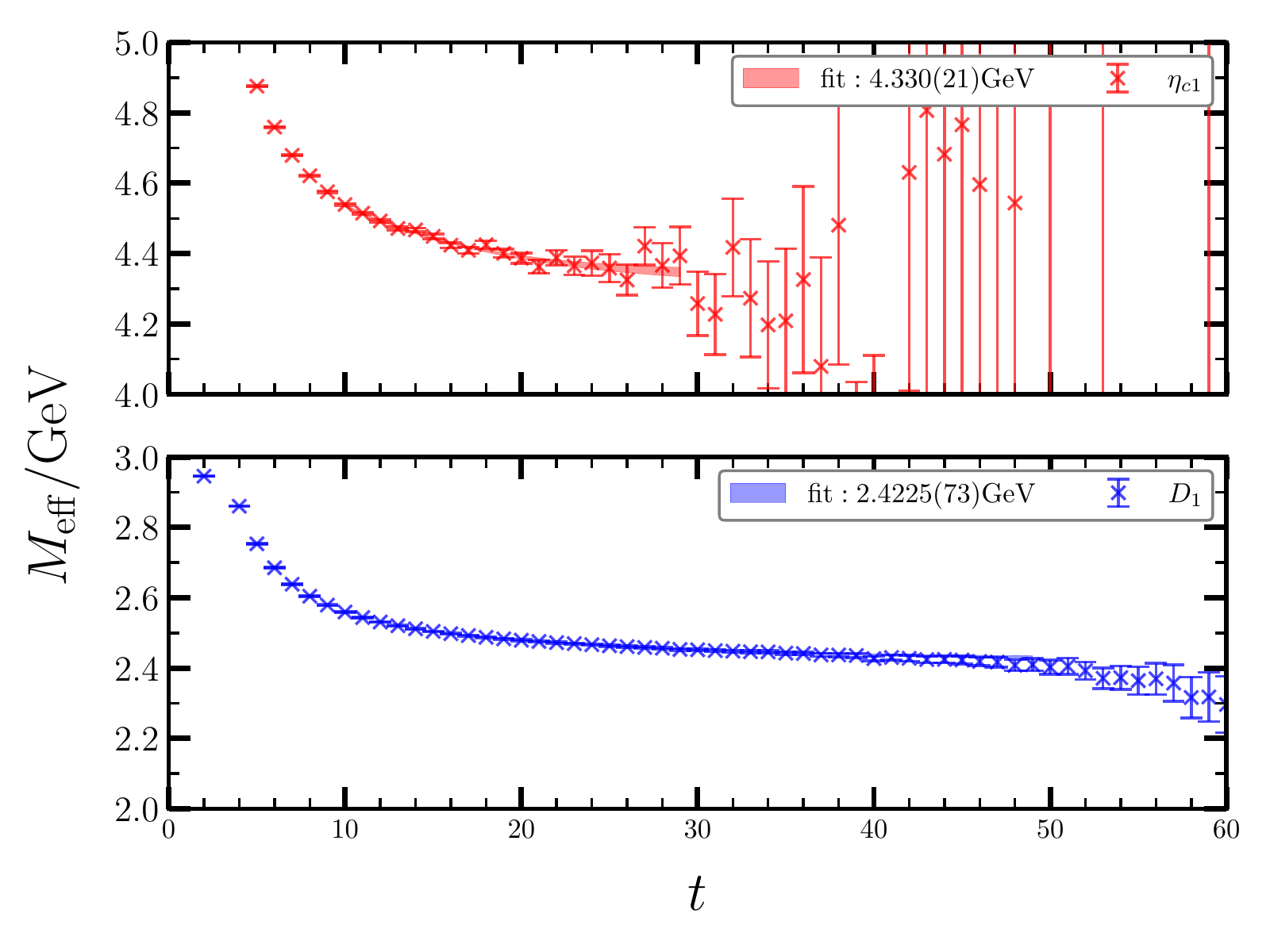}
    \caption{The effective masses of $1^{-+}$ charmoniumlike hybrid $\eta_{c1}$ (upper panel) and $D_1$ (lower panel) on L16. The data points are the lattice results, and the colored bands are the two-mass-term fits (Eq.~(\ref{eq:app-two-mass}) with the fitted parameters. The masses of $\eta_{c1}$ and $D_1$ on L16 are determined to be $m_{\eta_{c1}}=4.330(21)~\text{MeV}$ and $m_{D_1}=2.423(7)~\text{MeV}$, respectively.}
    \label{fig:effmass-etac-and-D1}
\end{figure}

\subsection{Fitting details}\label{sec:fitting-details}

As addressed in the main text, we use simple generic $\bar{q}\Gamma q$ operators for all the single mesons, all of which are ground states of specific quantum numbers. In the framework of the distillation methods, quark fields are Laplacian Heaviside smeared such that the coupling of these operators to excited states are suppressed. In order to determine the energy of each meson $X$ at a given momentum $k=|\vec{k}|$, we perform a two-mass fit to the correlation function $C_X(k,t)$ through the function form
\begin{eqnarray}\label{eq:app-two-mass}
    C_X(k,t)&=&W_1 \left(e^{-E_1(k)t}+e^{-E_1(k)(T-t)}\right)\nonumber\\
            &+&W_2 \left(e^{-E_2(k)t}+e^{-E_2(k)(T-t)}\right),
\end{eqnarray}
with the second term accounting for the contribution from higher states. Indeed, this function form presents accurately for all the single mesons. In Fig.~\ref{fig:effmass-etac-and-D1}, the data points are the effective masses of $\eta_{c1}$ (upper panel) and $D_1$ (lower panel) on L16 ensemble. The shaded band in each panel illustrates the fit time range and the fit quality. The correlations of other single mesons (mainly pseudoscalars and vectors) have much better signal-to-noise ratios (except for $\eta_{(2)}$). 

\begin{figure}[t]
	\includegraphics[width=.9\linewidth]{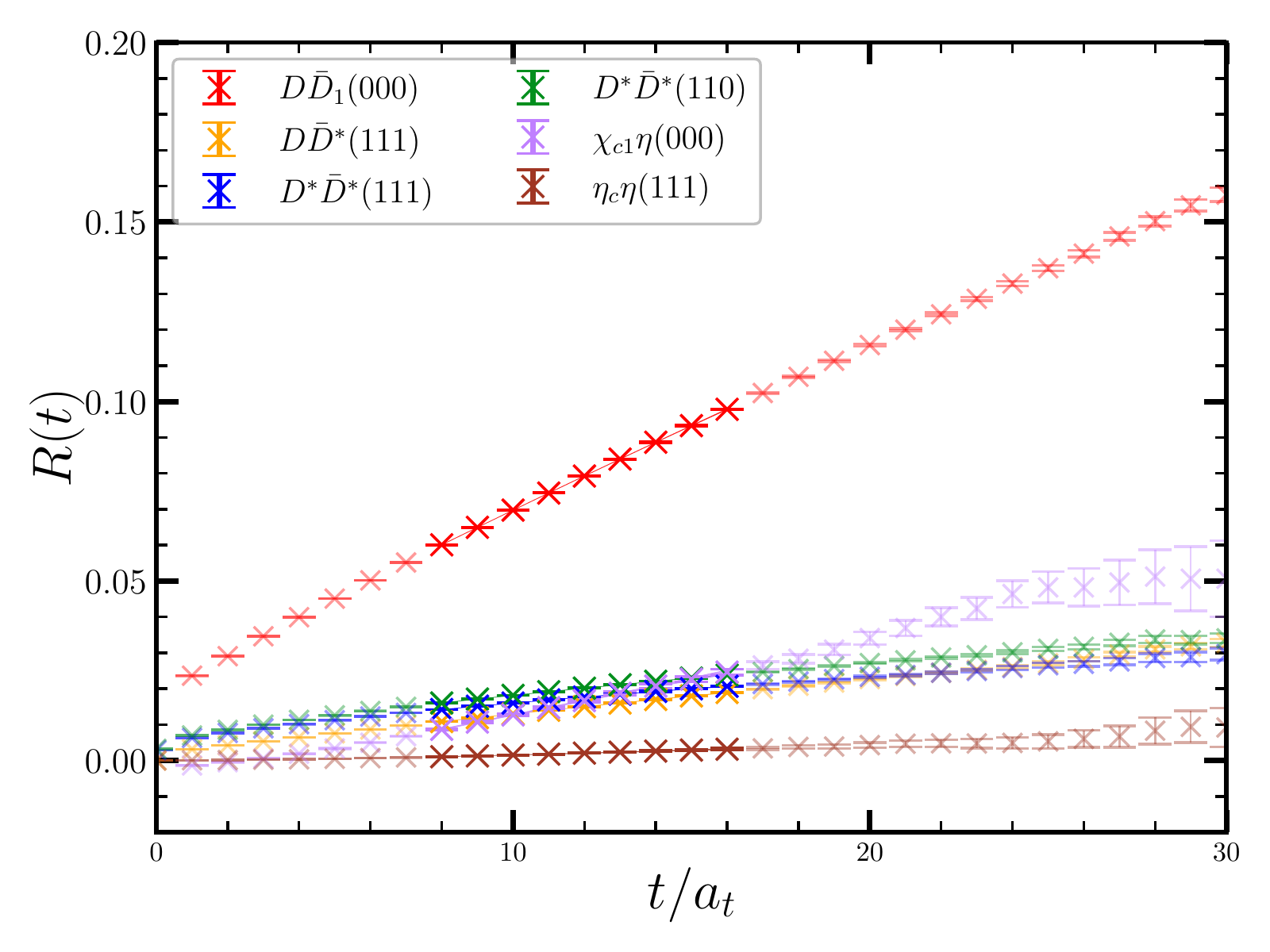}\\
	\includegraphics[width=.9\linewidth]{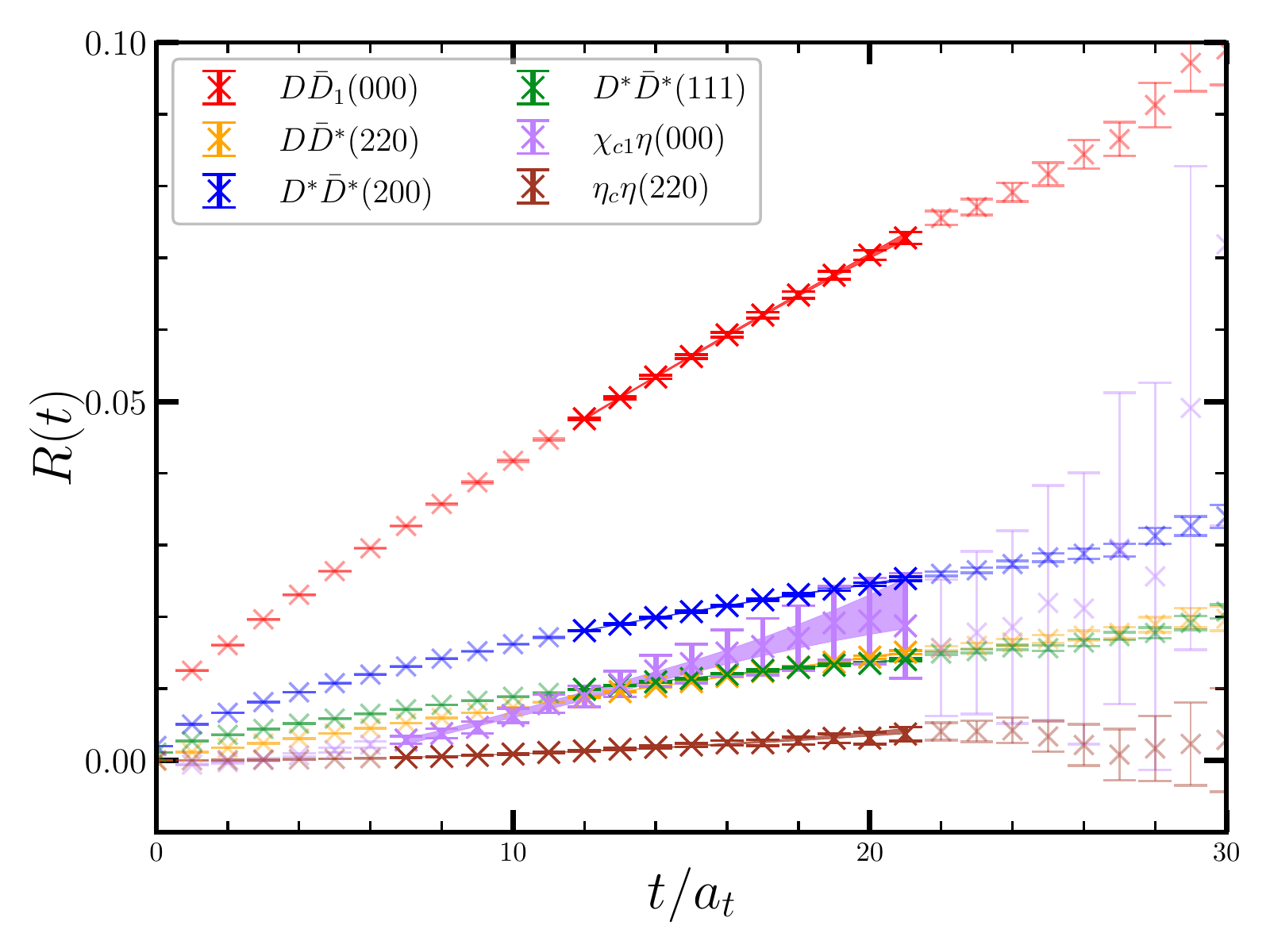}
	\caption{The ratio function $R_{AB}(\vec{k},t)$ on the two gauge ensembles L16 (the left panel) and L24 (the right panel) for $\eta_{c1}\to AB$. $AB$ refers to the decay modes $D_1\bar{D}$, $D^*\bar{D}$, $D^*\bar{D}^*$, $\chi_{c1}\eta_{(2)}$,  $\eta_c\eta_{(2)}$). The shaded bands illustrate the fit results and the fit window using Eq.~(\ref{eq:ratio-model-sm}).}
	\label{fig:slopes-sm} 
\end{figure}

For a two-body decay mode $AB$ of $\eta_{c1}$, the magnitude of the relative momentum $\vec{k}$ (denoted by the momentum mode $\hat k = (n_1,n_2,n_3)$) is chosen to give the value of $E_{AB}=E_A(\hat k)+E_B(\hat k)$
as close to $m_{\eta_{c1}}$ as possible. This is required for the derivation of Eq.~(\ref{eq:corr-HAB-sm}) and Eq.~(\ref{eq:ratio-sm}). For some decay modes, more than one momentum modes are assumed to check the momentum dependence of the transition amplitude and the effective coupling. Table~\ref{tab:amplitude} lists the $E_{AB}$ and $a_t \Delta=a_t(m_{\eta_{c1}}-E_{AB})$ for all the meson-meson modes involved. As shown in Eq.~(\ref{eq:ratio-sm}), the small
values $|a_t\Delta\lesssim| 0.01$ manifest that the derivation of $R_{AB}(\vec{k},t)$ from the linear behavior, depicted by $(a_t\Delta)^2 t^2/24$, is as small as 0.4\% even at $t=30$. 
Of course, the contamination from higher states will spoil the linear behavior, so we should fit $R_{AB}(\vec{k},t)$ using the function form of Eq.~(\ref{eq:ratio-sm}) in a time range where the contribution of higher states is unimportant. Figure~\ref{fig:slopes-sm} shows the $t$-behaviors of $R_{AB}(\vec{k},t)$ for all the decay channels concerned in this study. It is seen that the linear behavior does show up for each curve. The shaded curves illustrate the fit results and the fit time ranges using Eq.~(\ref{eq:ratio-model-sm}). The fitted values of $r_1$ and $r_3$ are also shown in Table~\ref{tab:amplitude}, where the absolute values of $r_3$ are smaller than those of $r_1$ by 
3-4 orders of magnitude. 
Consequently, the fitted values of $r_1$ are used to determine the final transition amplitude and the effective couplings.
\begin{figure}[t]
	\includegraphics[width=0.9\linewidth]{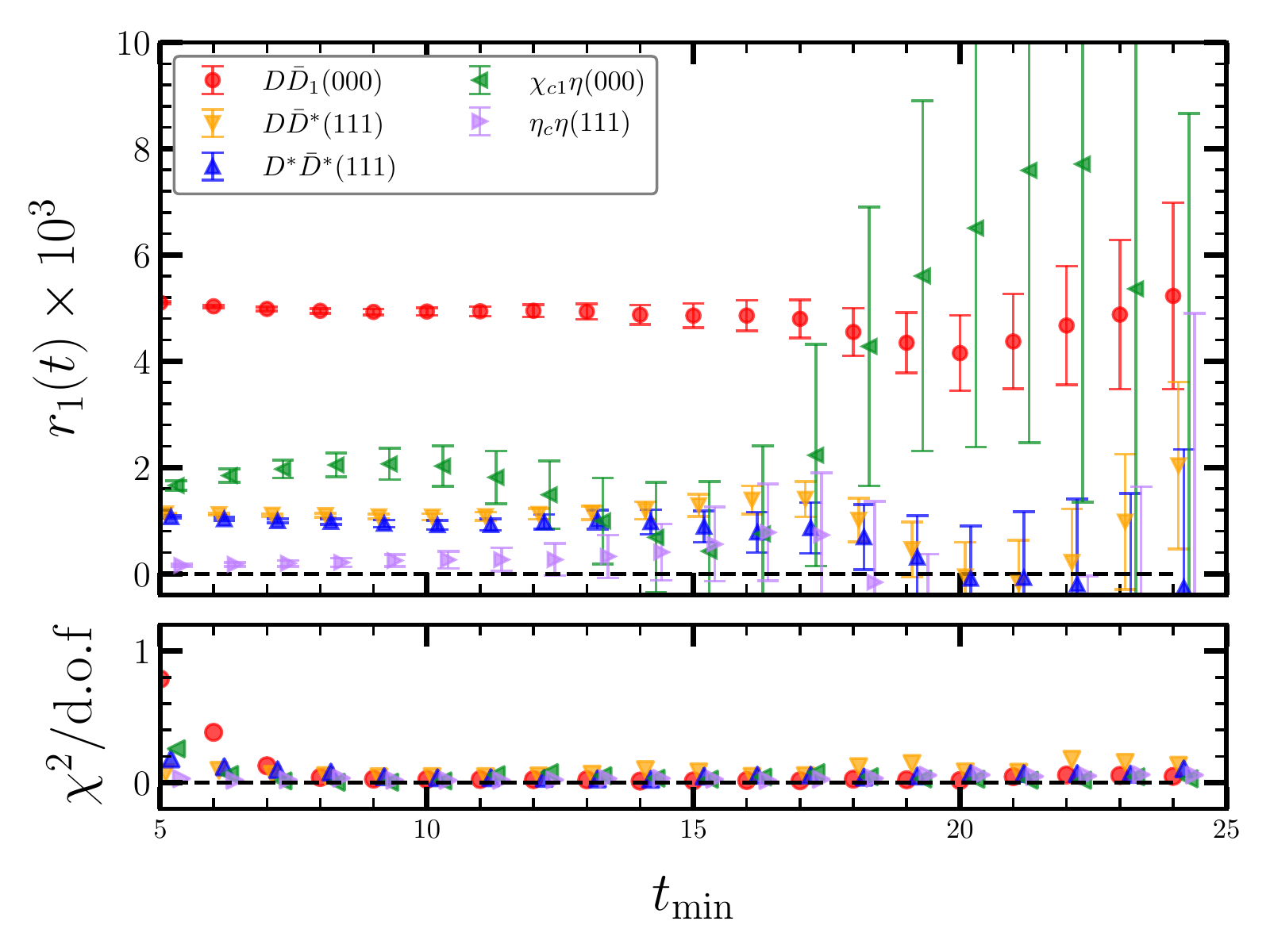}\\
	\includegraphics[width=0.9\linewidth]{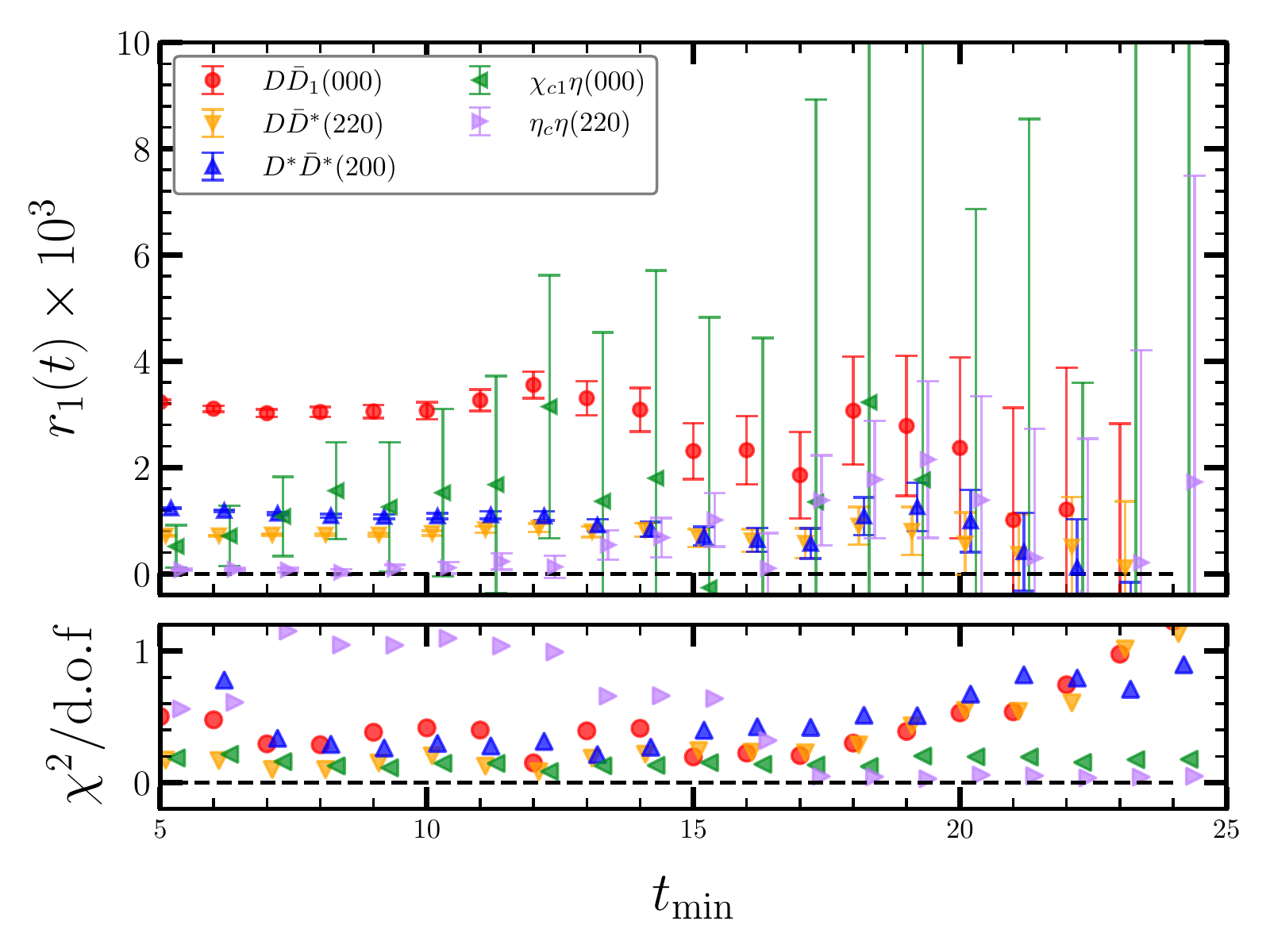}
	\caption{The fit stability of varying our fit window selections on our $r_1$ results. The horizontal axis, denoted $t_{\min}$, represents the case where we use the fit window $[t_{\min}, t_{\min} +10]$.}
	\label{fig:fit-stability}
\end{figure}
The fit stability of $R_{AB}(\vec{k},t)$ is also checked by varying the fit $t$ window. In doing so, we fix the length of the fit window to be 10 and conduct the fit in the time range $t\in [t_\text{min},t_\text{min}+10]$ by varying $t_\text{min}$ from 5 to 25. The $\chi^2/\text{d.o.f}$ values and the values of $r_1$ of the fits are illustrated in 
Fig.~\ref{fig:fit-stability}, where the left panels are the results of the L16 ensemble and the right ones are for the L24
ensemble. Obviously, the central values of $r_1$ for all the decay modes are stable when $t_\text{min}>7$, while the errors increase with the increasing of $t_\text{min}$. The smallness of the $\chi^2/\text{d.o.f}$ values for all the fits manifest the feasibility of the function form in Eq.~(\ref{eq:ratio-sm}). It is noted that the decay channels including isoscalar $\eta$ has worse signal at $t>25$ due to the contribution of the light annihilation diagrams. We take the fitted values of $r_1$ in the time ranges $[t_\text{min},t_\text{max}]=[8,18]$ as our final results, and take $\delta r_1=(r_1^\text{(max)}-r_1^\text{(min)})/2$ of the fits with $t_\text{min}\in[6,13]$ as a systematic error of $r_1$ and add it to the statistical error in a quadratic way.
\begin{table*}[t]
    \renewcommand\arraystretch{1.5}
    \caption{The energy $E_{AB}$, $a_t\Delta$ and the fitted $r_1$ for the six decay modes. For each mode at a specific relative momentum mode $(n_1n_2n_3)$, the value of $a_t\Delta=a_t(m_{\eta_{c1}}-E_{AB})$ is estimated by the central values of $m_{\eta_{c1}}$ and $E_{AB}$.}
    \label{tab:amplitude}
    \begin{ruledtabular}
        \scriptsize
        \begin{tabular}{llcccc}
                Mode   ($AB$)                   &  $E_{AB}$  (MeV)     &  $a_t\Delta$       & $r_0$      &$r_1$  $(\times 10^{-3})$     &      $r_3$  $(\times 10^{-7})$     \\\hline
             L16 &($m_{\eta_{c1}}=4330(21)~\text{MeV}$)         &           &               &               &                   \\\hline       
            $D\bar{D}_1(000)$           &  4304(7)      & $\sim$  0.004     &0.02077(38)        &4.95(5)         &    -5.1(1.3)      \\
            $D\bar{D}^*(111)$           &  4301(4)      & $\sim$  0.004     &0.00210(19)        &1.11(3)        &    -2.2(8)        \\
            $D^*\bar{D}^*(111)$         &  4405(10)     & $\sim$ -0.011     &0.00642(24)        &1.00(3)        &    -4.3(9)        \\
            $D^*\bar{D}^*(110)$         &  4288(4)      & $\sim$  0.006     &0.00702(25)        &1.15(4)        &    -4.1(9)        \\
            $\chi_{c1}\eta_{(2)}(000)$  &  4262(8)      & $\sim$  0.010     &-0.0076(18)        &2.04(26)        &    -0.9(6.8)      \\
            $\eta_c\eta_{(2)}(111)$     &  4347(39)     & $\sim$ -0.003     &-0.00057(39)       &0.20(6)        &    1.6(1.6)       \\\hline
             L24 &($m_{\eta_{c1}}=4328(68)~\text{MeV}$)         &           &               &               &                   \\\hline  
            $D\bar{D}_1(000)$           &  4310(10)     & $\sim$  0.003     & 0.0110(26)    &3.10(26)      &    -3.3(3.5)      \\
            $D\bar{D}^*(220)$           &  4361(5)      & $\sim$ -0.005     & -0.00027(69)  &0.78(7)        &    -1.07(97)      \\
            $D^*\bar{D}^*(200)$         &  4269(2)      & $\sim$  0.009     & 0.00598(87)   &1.05(9)        &    -3.1(1.2)      \\
            $D^*\bar{D}^*(111)$         &  4216(3)      & $\sim$  0.016     & 0.00234(62)   &0.67(7)        &    -2.50(83)      \\
            $\chi_{c1}\eta_{(2)}(000)$  &  4260(7)      & $\sim$  0.010     & -0.0057(27)   &1.18(38)       &    -3.0(9.0)        \\
            $\eta_c\eta_{(2)}(220)$     &  4304(25)     & $\sim$  0.004     & -0.00037(17)  &0.10(3)        &    2.49(77)      \\
        \end{tabular}
    \end{ruledtabular}
\end{table*}


The small but nonzero constant $r_0$ can be understood as follows. Assuming the correlation function $C_{AB,\eta_{c1}}^{33}(\vec{k},t)$ in Eq.~(\ref{eq:corr-veck-sm}) is contributed only from two orthogonal states $|\eta_{c1}\rangle$ and $|AB\rangle$ which are normalized as $\langle \eta_{c1}|\eta_{c1}\rangle =1$ and $\langle AB|AB\rangle=1$, one has 
\begin{eqnarray}
    C^{33}_{AB,\eta_{c1}}(\vec{k},t)&\equiv&\langle 0|\mathcal{O}^i_{AB}(\vec{k},t)\mathcal{O}^{j,\dagger}_{\eta_{c1}}(0)|0\rangle\nonumber\\
    &=&\langle 0|\mathcal{O}^3_{AB}(\vec{k},0) \hat{T}^t \mathcal{O}^{3,\dagger}_{\eta_{c1}}(0)|0\rangle, 
\end{eqnarray}
where $\hat{T}$ is the transfer matrix defined in Eq.~(\ref{eq:transfer-sm}) in the Hilbert space spanned by $|\eta_{c1}\rangle=\left(\begin{array}{c} 1\\ 0\end{array}\right)$ and $|AB\rangle=\left(\begin{array}{c} 0\\ 1\end{array}\right)$. For $(a_t\Delta)\ll1$ and $(a_t x_{AB})\ll 1$, one has 
\begin{widetext}
\begin{equation}
    \hat{T}^t=e^{-(a_t\hat{H}) t}=e^{-(a_t\bar{E})t} \left(
    \begin{array}{cc}
    e^{-(a_t\Delta)t/2 } & (a_t x_{AB})t\\
    (a_t x_{AB})t & e^{+(a_t\Delta)t/2}
    \end{array}\right)+\mathcal{O}(a_t^2 t^2 \Delta x_{AB}).
\end{equation}
Now considering 
\begin{eqnarray}
    \mathcal{O}_{\eta_{c1}}^{3,\dagger}|0\rangle &=& |\eta_{c1}\rangle\langle \eta_{c1}|\mathcal{O}_{\eta_{c1}}^{3,\dagger}|0\rangle + |AB\rangle\langle AB|\mathcal{O}_{\eta_{c1}}^{3,\dagger}|0\rangle \equiv \left(\begin{array}{c} Z_{\eta_{c1}}^{\eta_{c1}}\\Z_{\eta_{c1}}^{AB} \end{array}\right)\nonumber\\
    \mathcal{O}_{AB}^{3,\dagger}|0\rangle &=& |\eta_{c1}\rangle\langle \eta_{c1}|\mathcal{O}_{AB}^{3,\dagger}|0\rangle + |AB\rangle\langle AB|\mathcal{O}_{AB}^{3,\dagger}|0\rangle \equiv \left(\begin{array}{c} Z_{AB}^{\eta_{c1}}\\Z_{AB}^{AB} \end{array}\right),
\end{eqnarray}
the correlation function $C_{AB,\eta_{c1}}^{33}(\vec{k},t)$ can be expressed as 
\begin{equation}
    C_{AB,\eta_{c1}}^{33}(\vec{k},t)\approx Z_{AB}^{\eta_{c1}}\left(Z_{\eta_{c1}}^{\eta_{c1}}-Z_{\eta_{c1}}^{AB} (a_tx_{AB})t\right)e^{-m_{\eta_{c1}}t} + Z_{AB}^{AB}\left(Z^{AB}_{\eta_{c1}}-Z_{\eta_{c1}}^{\eta_{c1}} (a_t x_{AB})t\right)e^{-E_{AB}t}.
\end{equation}
\end{widetext}
If $\mathcal{O}_{\eta_{c1}}$ ($\mathcal{O}_{AB}$)does not couple to $|AB\rangle$ ($|\eta_{c1}\rangle$), namely, $Z_{\eta_{c1}}^{AB}=0$ and $Z_{AB}^{\eta_{c1}}=0$, then one can get the ratio function $R_{AB}(\vec{k},t)$ in Eq.~(\ref{eq:ratio-sm}). However, if $Z_{\eta_{c1}}^{AB}$ and $Z_{AB}^{\eta_{c1}}$ are nonzero but small, one can easily derive that 
\begin{eqnarray}
R_{AB}(\vec{k},t)&\approx& \left( \frac{Z_{AB}^{\eta_{c1}}}{Z_{AB}^{AB}}+\frac{Z^{AB}_{\eta_{c1}}}{Z_{\eta_{c1}}^{\eta_{c1}}}\right)-(a_t x_{AB})t+\ldots\nonumber\\
&\to& r_0+r_1 t+\ldots .
\end{eqnarray}
Thus the constant term $r_0\sim \left( \frac{Z_{AB}^{\eta_{c1}}}{Z_{AB}^{AB}}+\frac{Z^{AB}_{\eta_{c1}}}{Z_{\eta_{c1}}^{\eta_{c1}}}\right)$ serves as 
a measure of the validity of the assumption 
\begin{equation}\label{eq:approx-sm}
    \langle 0|\mathcal{O}_{\eta_{c1}}|AB\rangle\approx 0, ~~ \langle 0|\mathcal{O}_{AB}|\eta_{c1} \rangle\approx 0.
\end{equation}
We list $r_0$ of all the $AB$ modes in Table~\ref{tab:amplitude}, where one can see that all the values of $r_0$ are less than 0.02 and justifies assumptions in Eq.~(\ref{eq:approx-sm}).

\bibliography{ref}
\end{document}